\documentclass[fleqn,usenatbib]{mnras}

\usepackage{newtxtext,newtxmath}
 
\usepackage[T1]{fontenc}
\usepackage{ae,aecompl}

\usepackage{array}
\newcolumntype{?}{!{\vrule width 1.5pt}}
\usepackage{afterpage}

\usepackage{multirow}
\usepackage{graphicx}	
\usepackage[usenames]{color}
\usepackage{bm}		
\usepackage{booktabs,makecell}
\usepackage{adjustbox}
\usepackage{hyperref}
\usepackage{soul}
\usepackage{longtable}
\usepackage{comment}
\usepackage{xspace}

\usepackage{float}

\newcommand{\redmapper}{\textsc{redMaPPer}\xspace}
\newcommand{\redmagic}{\textsc{redMaGiC}\xspace}
\newcommand{\maglim}{\textsc{MagLim}\xspace}
\newcommand{\dnff}{\textsc{DNF}\xspace}

\usepackage[switch,pagewise]{lineno}

\usepackage{eso-pic}

\AddToShipoutPictureBG*{%
  \AtPageUpperLeft{%
    \hspace{0.75\paperwidth}%
    \raisebox{-3.5\baselineskip}{%
      \makebox[0pt][l]{\textnormal{DES }}
}}}%

\AddToShipoutPictureBG*{%
  \AtPageUpperLeft{%
    \hspace{0.75\paperwidth}%
    \raisebox{-4.5\baselineskip}{%
      \makebox[0pt][l]{\textnormal{FERMILAB-PUB-20-664-AE}}
}}}%

\makeatletter
\def\blfootnote{\xdef\@thefnmark{}\@footnotetext}
\makeatother

\title[DES Y3: Calibration of Lens Sample Redshift Distributions]{Dark Energy Survey Year 3 Results: Calibration of Lens Sample Redshift Distributions using Clustering Redshifts with BOSS/eBOSS}

\author[R. Cawthon et al.]{
\parbox{\textwidth}{
\Large
R.~Cawthon,$^{1,2*}$
J.~Elvin-Poole,$^{3,4}$
A.~Porredon,$^{3,5,6}$
M.~Crocce,$^{5,6}$
G.~Giannini,$^{7}$
M.~Gatti,$^{7}$
A.~J.~Ross,$^{3}$
E.~S.~Rykoff,$^{8,9}$
A.~Carnero~Rosell,$^{10,11}$
J.~DeRose,$^{12,13}$
S.~Lee,$^{14}$
M.~Rodriguez-Monroy,$^{15}$
A.~Amon,$^{8}$
K.~Bechtol,$^{1}$
J.~De~Vicente,$^{15}$
D.~Gruen,$^{16,8,9}$
R.~Morgan,$^{1}$
E.~Sanchez,$^{15}$
J.~Sanchez,$^{17}$
I.~Sevilla-Noarbe,$^{15}$
T.~M.~C.~Abbott,$^{18}$
M.~Aguena,$^{19,20}$
S.~Allam,$^{17}$
J.~Annis,$^{17}$
S.~Avila,$^{21}$
D.~Bacon,$^{22}$
E.~Bertin,$^{23,24}$
D.~Brooks,$^{26}$
D.~L.~Burke,$^{8,9}$
M.~Carrasco~Kind,$^{26,27}$
J.~Carretero,$^{7}$
F.~J.~Castander,$^{5,6}$
A.~Choi,$^{3}$
M.~Costanzi,$^{28,29}$
L.~N.~da Costa,$^{20,30}$
M.~E.~S.~Pereira,$^{31}$
K.~Dawson,$^{32}$
S.~Desai,$^{33}$
H.~T.~Diehl,$^{17}$
K.~Eckert,$^{34}$
S.~Everett,$^{13}$
I.~Ferrero,$^{35}$
P.~Fosalba,$^{5,6}$
J.~Frieman,$^{17,36}$
J.~Garc\'ia-Bellido,$^{21}$
E.~Gaztanaga,$^{5,6}$
R.~A.~Gruendl,$^{26,27}$
J.~Gschwend,$^{20,30}$
G.~Gutierrez,$^{17}$
S.~R.~Hinton,$^{37}$
D.~L.~Hollowood,$^{13}$
K.~Honscheid,$^{3,4}$
D.~Huterer,$^{31}$
D.~J.~James,$^{38}$
A.~G.~Kim,$^{39}$
J.-P.~Kneib,$^{40}$
K.~Kuehn,$^{41,42}$
N.~Kuropatkin,$^{17}$
O.~Lahav,$^{25}$
M.~Lima,$^{19,20}$
H.~Lin,$^{17}$
M.~A.~G.~Maia,$^{20,30}$
P.~Melchior,$^{43}$
F.~Menanteau,$^{26,27}$
R.~Miquel,$^{44,7}$
J.~J.~Mohr,$^{45,46}$
J.~Muir,$^{8}$
J.~Myles,$^{16,8,9}$
A.~Palmese,$^{17,36}$
S.~Pandey,$^{34}$
F.~Paz-Chinch\'{o}n,$^{47,27}$
W.~J.~Percival,$^{48}$
A.~A.~Plazas,$^{43}$
A.~Roodman,$^{8,9}$
G.~Rossi,$^{49}$
V.~Scarpine,$^{17}$
S.~Serrano,$^{5,6}$
M.~Smith,$^{50}$
M.~Soares-Santos,$^{31}$
E.~Suchyta,$^{51}$
M.~E.~C.~Swanson,$^{27}$
G.~Tarle,$^{31}$
C.~To,$^{16,8,9}$
M.~A.~Troxel,$^{14}$
and R.D.~Wilkinson$^{52}$
\begin{center} (DES Collaboration) \end{center}
}
\vspace{0.4cm}
\\
}

\date{Accepted XXX. Received YYY; in original form ZZZ}

\pubyear{2020}

\begin{document}
\label{firstpage}
\pagerange{\pageref{firstpage}--\pageref{lastpage}}
\maketitle

\begin{abstract}
We present clustering redshift measurements for Dark Energy Survey (DES) lens sample galaxies used in weak gravitational lensing and galaxy clustering studies. To perform this measurement, we cross-correlate with spectroscopic galaxies from the Baryon Acoustic Oscillation Survey (BOSS) and its extension, eBOSS. We validate our methodology in simulations, including a new technique to calibrate systematic errors due to the galaxy clustering bias, finding our method to be generally unbiased in calibrating the mean redshift. We apply our method to the data, and estimate the redshift distribution for eleven different photometrically-selected bins. We find general agreement between clustering redshift and photometric redshift estimates, with differences on the inferred mean redshift to be below $|\Delta z|=0.01$ in most of the bins. We also test a method to calibrate a width parameter for redshift distributions, which we found necessary to use for some of our samples. Our typical uncertainties on the mean redshift ranged from 0.003 to 0.008, while our uncertainties on the width ranged from 4 to 9\%. We discuss how these results calibrate the photometric redshift distributions used in companion DES Year 3 Results papers. 
\end{abstract}



\blfootnote{Affiliations are listed at the end of the paper.}
\blfootnote{*e-mail: rcawthon28@gmail.com}
\setcounter{footnote}{1}



\section{Introduction}
Large galaxy imaging surveys have proven to be an effective tool for understanding the cosmos. Optical surveys like the Dark Energy Survey (DES, \citealt{des05}), the Kilo-Degree Survey (KiDS, \citealt{kids}) and the Hyper Suprime-Cam (HSC, \citealt{hsc}) have shown the ability to catalog millions of galaxies and extrapolate cosmological information out to redshift, $z\sim1$, probing the structure and dynamics of the Universe in the past $\sim6$ billion years (\citealt{keypaper18}, \citealt{kidsshear}). Accompanying this work, \cite{y3-3x2ptkp} shows the latest analysis of structure of the Universe using galaxy clustering and weak lensing measurements of more than 100 million galaxies. In the future, surveys like the Vera Rubin Observatory Legacy Survey of Space and Time (LSST, \citealt{lsst19}) and Euclid (\citealt{euclid}) will extend such analyses to include billions of galaxies further back in time.


A critical component of these imaging surveys is the estimation of galaxy redshifts. Accurate redshift information is necessary for precise cosmological measurements of the growth of structure across time. However, large imaging surveys tend not to have spectroscopic capabilities. Instead, spectral information tends to be limited to magnitude-estimates in a few color bands. In DES, imaging data includes the $g,r,i,z$ and $Y$ bands. Estimating photometric redshifts (photo-$z$) is a topic with much literature (see \citealt{kidspz}, \citealt{despz18} and references therein). In these methods, a redshift estimate is extracted from these few color and magnitude measurements. These methods all require some form of testing on galaxies where photometric and spectroscopic measurements are taken. Despite much success with these methodologies, the best photometric redshift estimates in DES for particularly suitable samples of galaxies are thought to have uncertainties around $\sigma_{\text{z}} \approx 0.02$ for individual galaxies with many samples much more uncertain. These errors are orders of magnitude larger than typical spectroscopic redshift errors. One particular issue is that a systematic bias can emerge if the test samples of galaxies are not fully representative of the galaxies being studied (\citealt{rivera18}). This may happen, for example, from a difference in depth of the samples. Extrapolating from a few color-band measurements to a precise redshift remains a difficult problem.


In recent years, an alternative and complementary method of estimating redshifts of galaxies has developed. The approach, called `clustering redshifts' or `cross-correlation redshifts', computes an angular cross-correlation of the galaxy sample in question and a galaxy sample with known (spectroscopic) redshifts. This cross-correlation will contain a signal proportional to the redshift overlap of the two samples. The method is completely independent of photometry, not relying on the color-magnitude information at all (other than for initially binning the galaxies). Instead, it relies on gravity. Since galaxies cluster, objects near each other in angular coordinates are more likely to be near each other in radial separation, and thus redshift. While this spatial information will not be significantly informative on a galaxy by galaxy basis, it is very useful probabilistic information when trying to estimate the redshift distribution of thousands or millions of galaxies. 

The use of angular clustering to infer proximity in distance between two samples extends back to \cite{seldnerpeebles} and \cite{phillippsshanks}. The modern method of using that information for a rigorous estimate of a redshift distribution traces back to \cite{newman08}. It has since been developed theoretically and implemented on data in a number of papers including \cite{matthewsnewman}, \cite{mcquinnwhite}, \cite{menard13}, \cite{schmidt13},  \cite{choi16}, \cite{scottez16}, \cite{johnson17} \cite{unwise}, \cite{kidspz}, and \cite{kidswz}. In the DES Year-1 cosmology analysis (\citealt{keypaper18}), clustering redshifts of both lens and source galaxies were computed (\citealt{cawthon18}, \citealt{davis17}, \citealt{gatti18}). 


In this work, we present the clustering redshift estimates for DES `lens' galaxies used in the `Year-3' cosmological analyses (based on data from the first three years of DES observations). These galaxies are used as lenses for galaxy-galaxy lensing measurements, and for galaxy clustering measurements in the cosmology analysis in \cite{y3-3x2ptkp}. The lens galaxies and those measurements are analyzed in more detail in several related DES Year-3 analyses (\citealt{y3-2x2ptbiasmodelling}, \citealt{y3-2x2ptaltlensresults}, \citealt{y3-galaxyclustering}, \citealt{y3-gglensing}, \citealt{y3-2x2ptmagnification}). There are two samples of DES lens galaxies presented in these works: \redmagic and a magnitude-limited sample, called \maglim. \redmagic (\citealt{rozo16}) is an algorithm that finds luminous red galaxies (LRGs) by using the red sequence of galaxies (\citealt{gladdersyee}, \citealt{rykoff14}). This type of selection has been shown to give fairly small photometric redshift errors for the sample. A similar sample was used in the DES Year-1 analysis (\citealt{cawthon18}, \citealt{elvinpoole18}, \citealt{keypaper18}). The \maglim is a denser sample that goes to slightly higher redshifts, and is described in \cite{y3-2x2maglimforecast}. The \maglim sample is expected to have more uncertainty in its photo-$z$ estimates than \redmagic. The \redmagic and \maglim samples are split into five and six tomographic redshift bins respectively, selected by photo-$z$ estimates.


To calibrate the redshift distributions of these two samples, in each of their redshift bins, we cross-correlate them with spectroscopic samples of galaxies. For these spectroscopic samples, we use galaxies observed by the Sloan Digital Sky Survey (SDSS, \citealt{sdss1}, \citealt{sdss2}, \citealt{sdss3}, \citealt{sdss4}). Specifically, we use galaxies from the Baryon Oscillation Spectroscopic Survey (BOSS, \citealt{boss}), as was used in \cite{cawthon18}, as well as from the extended Baryon Oscillation Spectroscopic Survey (eBOSS, \citealt{eboss}). About $15\%$ of the DES Year-3 samples overlap BOSS and eBOSS.


Much of the methodology in this work is similar to that of \cite{cawthon18}. We briefly highlight the main differences in this analysis.

1. We use significantly larger datasets for DES and spectroscopic reference galaxies. In addition to the larger area of coverage for DES, we calibrate two lens samples (\redmagic and \maglim) while DES Year-1 results only used \redmagic. For spectroscopic samples, we are able to use more of BOSS due to the wider area of DES in Year-3. We also are able to use the eBOSS galaxy catalog which greatly improves the redshift coverage available, increasing the maximum redshift of our study from roughly $z=0.7$ to $z=1.15$. Due to both area and redshift coverage, the overall number of DES \redmagic galaxies and spectroscopic galaxies used in this work are each a factor of 10 larger than in \cite{cawthon18}. In addition, the \maglim sample is about 3.5 times larger than the \redmagic sample in the Year-3 studies.

2. While much of the methodology is the same, it is much more extensively tested in simulations. These tests were possible due to having simulated spectroscopic samples similar to BOSS and eBOSS catalogs. These tests give a more thorough estimate of the errors and uncertainties in the method. 


3. We introduce a novel step in correcting for the evolution of the galaxy clustering bias. The galaxy bias describes the relationship between the distribution of galaxies and of total matter. Change in this parameter with redshift within a single tomographic bin is known as a challenging systematic in the clustering redshifts method (see \citealt{kidswz} for a recent review of attempts to correct this effect). Auto-correlations of galaxies can in principle be used both for the photometric and spectroscopic samples as an estimate of the galaxy bias, which is then calibrated out of the clustering redshift estimate. Since the DES samples do not have spectroscopic redshifts, their auto-correlations are not only a function of the galaxy bias (and cosmology) but also the scatter in their true redshift distributions. In the DES Year-1 analysis in \cite{cawthon18}, this photo-$z$ scatter effect on the auto-correlations was calibrated from simulations. In our work, we calibrate this scatter effect with cross-correlations of the DES and spectroscopic samples on smaller redshift bins. The main advantage of this new step is that it is empirically driven, no longer assuming any information from simulations (although the step is tested along with all the others in simulations).

4. We test a few different `2-parameter fits', which effectively constrain both the mean redshift and the width of a distribution. In detail, the fits solve for a shift and a stretch of a photo-$z$ distribution to better match the clustering redshifts data. This procedure in practice is needed when the shapes of the two distributions mismatch, and a single shift parameter would not make them match well enough.

We also note that a clustering redshift measurement of the weak lensing `source' galaxies in \cite{y3-3x2ptkp} and companion papers is performed in \cite{y3-sourcewz}. That work has several similarities and differences in methodology compared to this work. One example is in constraining the galaxy bias evolution. For the lens sample, we have a generated random catalogue which samples the survey selection function. This gives us a greater ability to measure the galaxy bias evolution effects with auto-correlations. Since the source galaxies do not have such a catalog, \cite{y3-sourcewz} use a more agnostic model to account for galaxy bias evolution.

The structure of our paper is as follows. In Section \ref{sec:datasets}, we discuss the datasets used in this work. In Section \ref{sec:sims}, we describe the simulated datasets used for validating our methodology. In Section \ref{sec:methods}, we present our methodology for performing a clustering redshifts measurement and calibrating it to find a best fit shift, or shift and stretch parameters to be applied to a photometric estimate of the redshift distribution. In Section \ref{sec:simtests}, we validate our methodology in simulations and derive systematic uncertainties for different parts of the method, as well as test different methods for doing a 2-parameter fit. In Section \ref{sec:results}, we show our results, the clustering redshift measurements of each of the redshift bins of the two DES lens samples. In Section \ref{sec:magnification}, we calculate a theory prediction for magnification effects in our measurements, showing they are likely insignificant. In Section \ref{sec:summary}, we summarize our work.

\section{Datasets}
\label{sec:datasets}

In this section, we describe the datasets used for the spectroscopic reference galaxies and the photometric DES galaxies that we wish to calibrate. The \redmagic and \maglim samples are derived from the `Y3 Gold catalog' (\citealt{y3-gold}) which contains galaxies found in the first three years of DES data. The Gold catalog covers the full DES footprint of nearly $5000$ $\text{deg}^2$, and contains around 388 million objects. The two samples are used for cosmological analyses in \cite{y3-3x2ptkp} and are described in detail in \cite{y3-galaxyclustering} and \cite{y3-2x2maglimforecast}. We repeat some of the main information about each sample here. This work only uses the part of the DES catalogs that overlaps the sky area of BOSS or eBOSS galaxies, about $860 \ \text{deg}^2$, with slightly less overlap at higher redshifts  (see Table \ref{table:spec_ngals}). Masks are also derived from the Gold catalog as well as random galaxy catalogs which reflect the survey selection efficiency at different points. After masks are applied, the effective DES area in our study is $632 \ \text{deg}^2$. We note that in Appendix \ref{sec:fluxlim}, we describe and analyze a third sample, called `flux-limited' which is not used in the cosmology analyses.

\subsection{Dark Energy Survey \redmagic}
\label{sec:des}
To create the \redmagic sample, the cluster-finding algorithm \redmapper (\citealt{rykoff14}) is run on the Gold catalog to calibrate the red sequence of galaxies (\citealt{gladdersyee}). The \redmagic algorithm (\citealt{rozo16}) then selects luminous red galaxies with colors that fit with the red sequence template. This fitting also estimates a redshift probability distribution function for each LRG. The \redmagic algorithm further tunes the color selection threshold to produce a constant comoving density, which is expected for passively evolving red galaxies (\citealt{rozo16}). \redmagic galaxy catalogs were similarly used for DES Year-1 analyses (\citealt{keypaper18}, \citealt{elvinpoole18}, \citealt{cawthon18}).

The \redmagic algorithm selects galaxies above a given luminosity threshold. For the Year-3 lens samples, thresholds of either 0.5 $L_*$ or 1.0 $L_*$ were used for the different redshift bins (see Table \ref{table:redmagic_ngals}). The reference luminosity, $L_*$, comes from a model (\citealt{bruzualcharlot}) for a single star-formation burst at $z=3$, as described in \cite{rykoff14}. For the reference luminosities 0.5 $L_*$ and 1.0 $L_*$, the comoving densities produced by the \redmagic algorithm are $\bar{n}=10^{-3} \ \text{and} \ 4*10^{-4}$ galaxies/$(h^{-1} \text{Mpc})^3$ respectively, with $h$ being the reduced Hubble constant (\citealt{y3-galaxyclustering}). The \redmagic galaxies are split into tomographic bins by the mean redshift of each galaxy's redshift probability distribution function produced by the \redmagic algorithm. We show the number of galaxies used in this analysis (covering the $632 \ \text{deg}^2$ of overlap with BOSS) for each tomographic bin in Table \ref{table:redmagic_ngals}.

We also apply weights to \redmagic galaxies as described in \cite{y3-galaxyclustering}. These weights are selected based on survey properties like seeing and sky brightness for each of the observed galaxies. The weights are chosen to minimize the impacts of these survey properties on galaxy clustering measurements.

\begin{table}
\begin{center}
    \begin{tabular}{|c|c|c|c|}
      \hline
      \multicolumn{4}{|c|}{DES \redmagic Samples} \\
      \hline
      Redshift Bin & $L/L_{*}$ & $n_{\text{gal}} \ [\text{arcmin}^{-2}]$ & $N_{\text{gal}}$ \\
      \hline
      1: $z_{\text{ph}} \in [0.15,0.35]$ & 0.5 & 0.027 & 61586  \\   
      \hline
      2: $z_{\text{ph}} \in [0.35,0.5]$ & 0.5 & 0.049 & 110586  \\   
      \hline
      3: $z_{\text{ph}} \in [0.5,0.65]$ & 0.5 & 0.075 & 170102  \\   
      \hline
      4: $z_{\text{ph}} \in [0.65,0.8]$ & 1.0 & 0.038 & 86767  \\   
      \hline
      5: $z_{\text{ph}} \in [0.8,0.9]$ & 1.0 & 0.032 & 72833  \\   
      \hline
    \end{tabular}
  \caption{\redmagic galaxies used in this work.}
  \label{table:redmagic_ngals}
\end{center}
\end{table}

\begin{table}
\begin{center}
    \begin{tabular}{|c|c|c|}
      \hline
      \multicolumn{3}{|c|}{DES \maglim Samples} \\
      \hline
      Redshift Bin & $n_{\text{gal}} \  [\text{arcmin}^{-2}]$ & $N_{\text{gal}}$ \\
      \hline
      1: $z_{\text{ph}} \in [0.2,0.4]$ & 0.154 & 349673  \\   
      \hline
      2: $z_{\text{ph}} \in [0.4,0.55]$ & 0.115 & 260671  \\   
      \hline
      3: $z_{\text{ph}} \in [0.55,0.7]$ & 0.115 & 262468  \\   
      \hline
      4: $z_{\text{ph}} \in [0.7,0.85]$ & 0.154 & 349996  \\   
      \hline
      5: $z_{\text{ph}} \in [0.85,0.95]$ & 0.117 & 266750  \\   
      \hline
      6: $z_{\text{ph}} \in [0.95,1.05]$ & 0.113 & 257139  \\   
      \hline
    \end{tabular}
  \caption{\maglim galaxies used in this work.}
  \label{table:maglim_ngals}
  \end{center}
\end{table}

\subsection{DES \maglim sample}
\label{sec:maglim}
The DES \maglim samples are described in \cite{y3-2x2maglimforecast}. They are created using a redshift-dependent magnitude cut, with the redshift estimate for each galaxy coming from the \dnff (\citealt{dnfer}) photometric redshift algorithm. This redshift dependence tends to eliminate faint, low-redshift galaxies from entering the sample (as verified in Appendix \ref{sec:fluxlim}). The creation of this sample was motivated by a significantly larger number density than \redmagic. However, photo-$z$ error estimates were expected to be larger, making the calibration of photo-$z$ biases in this work essential.

In \cite{y3-2x2maglimforecast}, a Fisher forecast is run to find the best magnitude cuts for the DES cosmology analyses, with different cuts trading off number density and larger photo-$z$ scatter. From that work, the optimal redshift-dependent cut is selecting on $i$-band magnitude, $i <4 z_{\text{phot}}+18$. Bright galaxies with $i<17.5$ are also removed. These \maglim galaxies are split into tomographic bins by the mean redshift of the redshift probably distribution function given by \dnff. Notably, the \maglim sample extends to slightly higher redshifts than \redmagic. The numbers of galaxies used in this work for each tomographic bin (again reflecting only the overlapping galaxies with BOSS) are shown in Table \ref{table:maglim_ngals}. We again also apply survey property weights as described in \cite{y3-galaxyclustering}.

\begin{figure}
\begin{center}
\includegraphics[width=0.5 \textwidth]{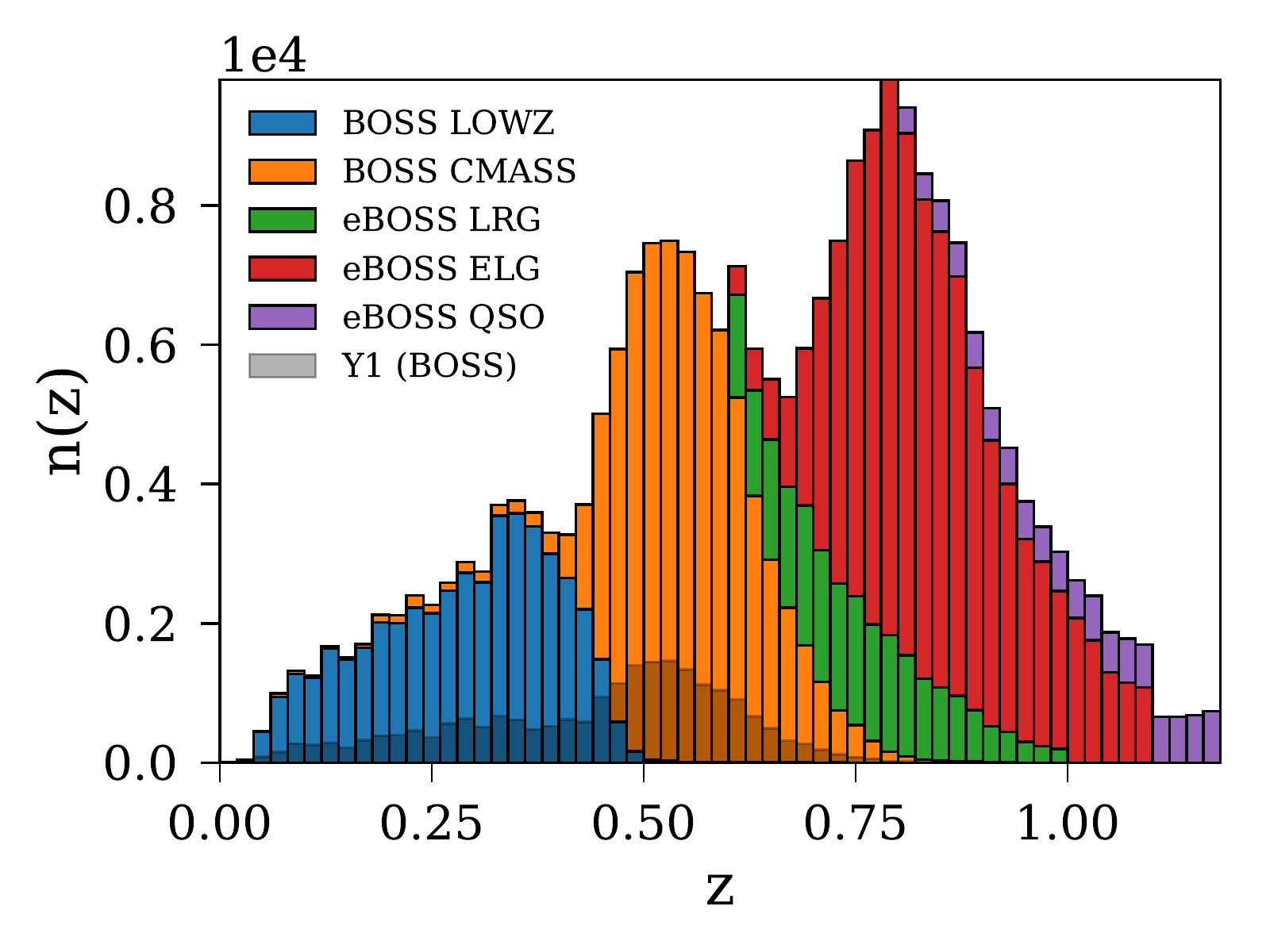}
\caption{The BOSS/eBOSS $n(z)$ used in this work as reference samples. In shaded outline, we show the BOSS $n(z)$ used in the Year-1 analysis of \protect\cite{cawthon18}. This work uses about a factor of 10 more reference galaxies overall. The numbers of galaxies from each BOSS/eBOSS catalog are shown in Table \ref{table:spec_ngals}.}
\label{fig:ebossnz}
\end{center}
\end{figure}

\subsection{Baryon Oscillation Spectroscopic Survey Galaxies (SDSS DR12)}
\label{sec:boss}

Our first source of reference galaxies comes from  catalogs created by the Baryon Oscillation Spectroscopic Survey from SDSS Data Release 12 (DR12, \citealt{dr12}). We use their LOWZ and CMASS galaxy and random catalogs described in \cite{reid16}. For $z>0.6$, we use a joint catalog of CMASS galaxies and eBOSS LRG galaxies created by eBOSS to prevent double counting of galaxies (see \citealt{ross20}). Our spectroscopic tracers at $z<0.6$ are solely from the BOSS samples. These samples were also used for clustering redshifts in DES Y1 cosmology (\citealt{keypaper18}, \citealt{cawthon18}). These catalogs were optimized for clustering in order to measure the Baryon Acoustic Oscillation (BAO) signal (\citealt{bossbao}), but their wide, uniform coverage of the sky makes them one of the best spectroscopic datasets for clustering redshifts.

\subsection{eBOSS (SDSS DR16)}
\label{sec:eboss}
We also use spectroscopic galaxies from the extended Baryon Oscillation Spectroscopic Survey. The galaxies are part of the SDSS Data Release 16 (DR16, \citealt{dr16}). We use the large-scale structure (LSS) catalogs of emission line galaxies (ELGs), luminous red galaxies (LRGs) and quasi-stellar objects (QSOs). The creation of the ELG catalogs is described in \cite{raichoor20} and the LRG and QSO catalogs are described in \cite{ross20}. The catalogs were provided to DES before being made public for clustering redshifts usage by agreement between DES and eBOSS. 

The target selection for the eBOSS ELG sample is described in \cite{raichoor17}. The sample selection is based on observations from the Dark Energy Camera Legacy Survey (DECaLS, \citealt{decals}) with color and magnitude cuts to yield strong [OII] emitters in the redshift range of $0.6<z<1.1$. The LRG sample selection is described in \cite{prakash16}. The LRGs were selected using color and magnitude cuts on objects found in SDSS and Wide-Field Infrared Survey Explorer (WISE, \citealt{wise}) photometry. The LRG sample primarily spans the range $0.6<z<1.0$. The LRGs are combined with the BOSS CMASS sample since there are duplicate objects. The QSO sample selection is described in \cite{myers15}. The QSO sample spans from $0.9<z<2.2$, although we only use up to $z=1.18$ for clustering redshift measurements due to low density of objects at higher redshifts for both DES and reference samples. The target selection used photometric observations from SDSS as well as WISE.

The details of creating large-scale structure datasets from these samples are described in \cite{raichoor20} and \cite{ross20}. We use the weights (given by $w_{\text{tot}}$ in \citealt{ross20}), and random points associated with these catalogs to account for the survey selection function. We also use the combined LRG catalog using eBOSS galaxies as well as $z>0.6$ BOSS CMASS galaxies, as described in \cite{ross20}. We show the total number of reference galaxies used in this work by their catalog source in Figure \ref{fig:ebossnz} and Table \ref{table:spec_ngals}. In our measurements, we combine all these catalogs into a single sample. 

\begin{table}
\begin{center}
    \begin{tabular}{|c|c|c|c|}
      \hline
      \multicolumn{4}{|c|}{Spectroscopic Samples} \\
      \hline
      Name & Redshifts & $N_{\text{gal}}$ & Area \\
      \hline
      LOWZ (BOSS) & $z \sim [0.0,0.5]$ & 45671 & $\sim860 \  \text{deg}^2$   \\   
      \hline
      CMASS (BOSS) & $z \sim [0.35,0.8]$ & 74186 & $\sim860 \  \text{deg}^2$  \\ 
      \hline
      LRG (eBOSS) & $z \in [0.6,1.0]$ & 24404 & $\sim700 \  \text{deg}^2$  \\   
      \hline
      ELG (eBOSS) & $z \in [0.6,1.1]$ & 89967 & $\sim620 \  \text{deg}^2$  \\   
      \hline
      QSO (eBOSS) & $z \in [0.8,1.18]$ & 10502 & $\sim700 \  \text{deg}^2$  \\   
      \hline
    \end{tabular}
  \caption{Spectroscopic samples used as the reference galaxies for clustering redshifts in this work. We show the approximate redshift range of the BOSS samples used. In contrast, the eBOSS catalogs each have set redshift boundaries.}
  \label{table:spec_ngals}
  \end{center}
\end{table}

\section{Simulated Datasets}
\label{sec:sims}

Our work, as well as many of the other accompanying papers related to \cite{y3-3x2ptkp}, make use of the Buzzard simulations (\citealt{y3-simvalidation}, \citealt{buzzard}). Buzzard simulates a dark matter only universe, which is then populated with galaxies by the ADDGALS algorithm (\citealt{addgals}). ADDGALS is calibrated by a series of algorithms, many of which are fit empirically to galaxy distributions (in terms of luminosity, clustering, abundance etc.) of SDSS galaxies (e.g., sub-halo abundance matching fits from \citealt{lehmann17}). The resulting galaxy catalogs are then sampled similarly to how DES creates its cosmological datasets. For the samples used in this work, this specifically means running the \redmapper and \redmagic algorithms on Buzzard to create a simulated \redmagic catalog, and using the color and magnitude cuts from the \maglim sample to create a simulated version of it. The Buzzard simulations used for DES Year-3 analyses are described in more detail in \cite{y3-simvalidation}. In that work, the simulated DES datasets are shown to replicate well galaxy properties and cosmological measurements from data.

\subsection{Simulated DES \redmagic}
As described in \cite{y3-simvalidation}, the \redmagic algorithm is run on Buzzard galaxies similar to the procedure on data. In particular, color-dependent clustering was improved for Year-3 Buzzard to better match the \redmagic-selected galaxies in data. The same redshift and $L_*$ cuts applied on the data are applied to get the simulated samples.

\subsection{Simulated DES \maglim sample}

For \maglim galaxies in Buzzard, we use a similar redshift-dependent magnitude cut as is done on data. For the tests in this work, a slightly older version of the \maglim cuts was used to generate the sample. This version selected galaxies with $i$-band magnitude, $i <4.28 z_{\text{phot}}+18$. It also cut out bright galaxies with $i<17.5$. The slight differences from the final \maglim cuts on data should not change the efficacy of our clustering redshift method, which is what the simulations are used to check. We do not use any information on, e.g., galaxy bias or photo-$z$ scatter from these simulated samples in our measurements. The redshift bins are selected in the same way as on data.

\subsection{Simulated BOSS (CMASS) sample}
To simulate the BOSS CMASS sample in Buzzard, we use the DMASS algorithm described in \cite{dmass}. The goal of the DMASS algorithm was to create a CMASS-like sample of galaxies from DES samples of galaxies. Since the properties of CMASS galaxies have been well characterized, a large CMASS-like sample in DES would be useful for several studies.

In \cite{dmass}, the DMASS algorithm is trained on the overlapping area of DES and BOSS to derive a Bayesian model based on galaxy colors and magnitudes for any DES galaxy to be CMASS-like. For our work, this algorithm is used on the Buzzard simulated DES galaxies. Each galaxy is given a CMASS-like probability. We then take one random draw based on these probabilities to define our simulated CMASS sample.

\subsection{Simulated eBOSS (ELG) sample}
To simulate the eBOSS ELG sample in Buzzard, we use the magnitude and color cuts used for target selection in \cite{raichoor17} for the South Galactic Cap (SGC) sample. Since these targets were found by the Dark Energy Camera (DECam), the targeting magnitudes are in the DECam filter bands.

\begin{figure}
\begin{center}
\includegraphics[width=0.5 \textwidth]{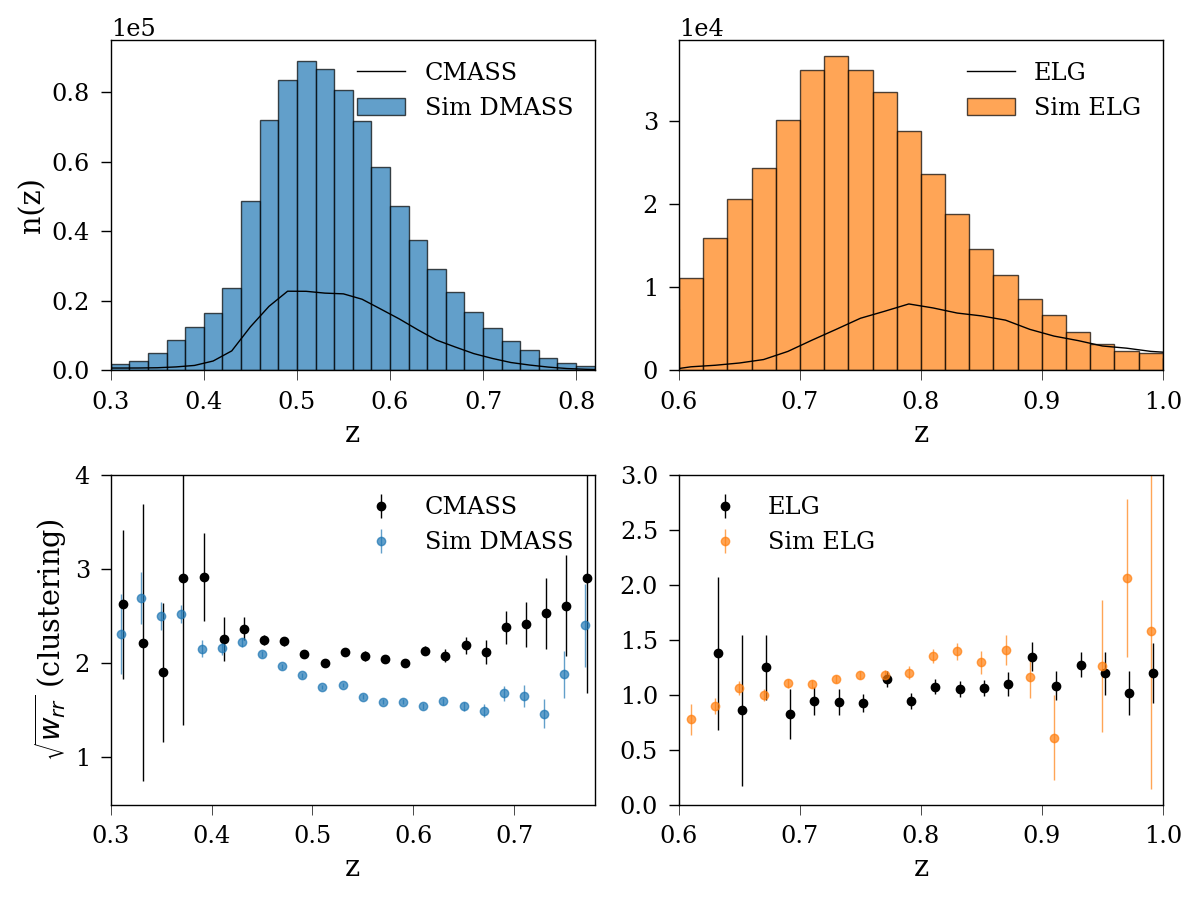}
\end{center}
\caption{Comparison of simulated and real spectoscopic datasets. Shown are BOSS CMASS (South), and DMASS, an algorithm run on simulations. Also shown is eBOSS ELG (South Galactic Cap) and a simulated ELG sample. The top row shows the redshift distributions. The bottom row shows the square root of a weighted auto-correlation (Equation \ref{autocorrelation}). The simulated samples go over the full DES 5000 $\text{deg}^2$, so they are larger than the real datasets.}
\label{fig:sim4plot}
\end{figure}

We show comparisons of the simulated BOSS/eBOSS samples to their real counterparts in Figure \ref{fig:sim4plot}. There is relative agreement in both redshift distribution and amount of clustering. We note that perfect agreement is not necessary, since the clustering and redshift distribution of the BOSS/eBOSS samples are well measured on the data, and there is no explicit reason the method's accuracy should depend strongly on redshift or galaxy bias. The relative agreement should be sufficient to validate the methodology. We do investigate the method's dependence on the number of galaxies in Appendix \ref{sec:densitydependenceappendix}.

\section{Methods}
\label{sec:methods}

We now lay out our methodology for the clustering redshift measurement. A number of different redshift distributions, binning schemes and correlation functions will be mentioned in this Section. To aid the reader, a summary of various terms is shown in Table \ref{table:terms}.

\begin{table}
\begin{center}
    \begin{tabular}{|>{\centering\arraybackslash}m{3.7cm}|>{\centering\arraybackslash}m{3.7cm}|}
      \hline
      \multicolumn{2}{|c|}{Definitions Guide} \\
      \hline
      \multicolumn{2}{|c|}{Redshift Distributions} \\
      \hline
      $n_{\text{u},i}(z)$ & The true redshift distribution of an unknown sample in photometric redshift bin $i$.  \\
      \hline
      $n_{\text{u},j}(z)$ & The true redshift distribution of a sample binned by photometric redshift in micro-bin $j$ \\
      \hline
      $n_{\text{spec},j}(z)$ & The true redshift distribution of a sample binned by spectroscopic (true) redshift in micro-bin $j$ \\
      \hline
      $n^{\text{pz},i}(z)$ & The photometric redshift distribution of a photometric redshift bin, $i$. \\   
      \hline
      \multicolumn{2}{|c|}{Binning Schemes} \\
      \hline
      Photometric bins, $i$ & The main target bins used for cosmology. Range in size from $dz=0.1-0.2$. \\   
      \hline
      Micro-bins, $j$ & The reference sample bins of size $dz=0.02$. Unknown and reference samples use these bins for auto-correlations. \\   
      \hline
      Nano-bins & Bins of size dz=0.005-0.01. These are used for computing the width of $n_{j}(z)$ in Equation \ref{stnddev}. \\   
      \hline
      \multicolumn{2}{|c|}{Correlation Functions} \\
      \hline
      $\bar{w}_{\text{ur}}$ & Weighted cross-correlation between an unknown (photometric) and reference (spectroscopic) galaxy samples. \\   
      \hline
      $\bar{w}_{\text{rr}}$ & Weighted auto-correlation of a reference sample. \\   
      \hline
      $\bar{w}_{\text{uu,pz}}$ & Weighted auto-correlation of an unknown sample which is binned by photometric redshift. \\   
      \hline
      $\bar{w}_{\text{uu,spec}}$ & Weighted auto-correlation of an unknown sample which is binned by spectroscopic redshift (typically not possible with data). \\   
      \hline
    \end{tabular}
  \caption{Definitions for various redshift distributions, binning schemes and correlation functions referred to in Section \ref{sec:methods}.}
  \label{table:terms}
\end{center}
\end{table}

\subsection{Unknown and Reference Correlation Measurement}
\label{sec:31}

The clustering redshift methodology involves a cross-correlation of two samples, an `unknown' sample with undetermined redshifts, and a `reference' sample with known redshifts. For this work, the unknown samples will be the DES samples (\redmagic and \maglim) and the reference sample will be the combined BOSS/eBOSS spectroscopic dataset.

We use a cross-correlation version of the Landy-Szalay estimator \citep{landyszalay} over physical scales $r$:

\begin{equation}
\label{landyszalay}
w(r)=\frac{D_1 D_2(r) - D_1 R_2(r) - D_2 R_1(r)+R_1 R_2(r)}{R_1 R_2(r)}
\end{equation}

\noindent where $w(r)$ is the excess probability of finding a pair of galaxies $r$ distance away compared to a random sample, $D$ signifies a dataset of galaxies, and $R$ signifies a random distribution of galaxies, and e.g., $D_1 D_2$ is the number of pairs between the two datasets separated by comoving length scale $r$. The length scale is set by $r=\theta \chi(z)$, where $\theta$ is the observed angle between the two galaxies and $\chi(z)$ is the comoving distance calculated using the Planck 2015 cosmology \cite{planck15}. The redshifts will be set by the center of the reference sample bins. For all of our measurements, we will use a weighted averaged estimate of $w(r)$ over a range of $r$ values, $\bar{w}_{12}$:

\begin{equation}
\label{lsequation}
\bar{w}_{12}=\int_{r_{\text{min}}}^{r_{\text{max}}} r^{-1} w(r)  dr.
\end{equation}

Unless otherwise stated, we use eight bins between $r_{\text{min}}=0.5$ Mpc and $r_{\text{max}}=1.5$ Mpc. These parameters, as well as the weighting by $r^{-1}$, were first shown effective for clustering redshifts in \cite{schmidt13} and were used in the DES Year-1 analyses: \cite{cawthon18}, \cite{davis17} and \cite{gatti18}. These comoving scales are smaller than the scales used for related cosmological galaxy clustering studies in \cite{y3-3x2ptkp} and others to reduce covariance of the measurements. For all of the following weighted cross- and auto-correlations given by Equation \ref{lsequation}, statistical errors are measured by 100 jackknife resamplings.

Our weighted cross-correlation, $\bar{w}$ of the unknown (u) and reference (r) samples should go as:

\begin{equation}
 \bar{w}_{\text{ur}} = \int_{z_{\text{min}}}^{z_{\text{max}}}  n_{\text{u}}(z) n_{\text{r}}(z) b_{\text{u}}(z)b_{\text{r}}(z)\bar{w}_{\text{mm}}(z) dz
\label{crosscorr}
\end{equation}

\noindent where $n_{\text{u}}$ and $n_{\text{r}}$ are the normalized redshift distributions of the unknown and reference galaxy samples, $b_{\text{u}}$ and $b_{\text{r}}$ are the galaxy biases of the two samples, and $\bar{w}_{\text{mm}}$ is the weighted cross-correlation of the total (primarily dark) matter distribution.

We now introduce two redshift binning schemes important for our work. The goal is to derive correct mean redshifts for the photometric redshift-binned DES samples to be used in the DES cosmology analyses. There are five bins for \redmagic, and six bins for the \maglim sample. We will call these the photometric bins and they will be signified by $i$. These bins are typically $dz=0.1-0.2$ in size. To obtain a measurement of $n_{\text{u},i}(z)$ relevant for these photometric bins, we need to measure on thinner bin widths. These thinner bins will be of size $dz=0.02$. We will call them the micro-bins and they will be signified by $j$. Our spectroscopic reference samples will always be binned in these smaller $dz=0.02$ bins. We will refer to the centers of these micro-bins with $z_j$. Our goal is to measure the photometric redshift sample's redshift distribution, $n_{\text{u},i}$, in each of the micro-bins at $z_j$. The characters u and r will always refer to samples, and the characters $i$ and $j$ will always refer to bins. In one micro-bin, we know the exact number of galaxies in the reference sample (since it has spectroscopic redshifts). Going from Equation \ref{crosscorr}, our estimate for $n_{\text{u},i}$ at a micro-bin centered at $z_j$ is:

\begin{equation}
\label{nequation1}
n_{\text{u},i}(z_j) \propto  \bar{w}_{\text{ur}}(z_j)  \frac{1}{{b}_{\text{u}}(z_j)} \frac{1}{{b}_{\text{r}}(z_j)}
  \frac{1}{\bar{w}_{\text{mm}}(z_j)}
\end{equation}

\noindent where $n_{\text{u},i}(z_j)$ is the desired quantity of the number of galaxies in unknown (photometric) sample $i$ in the micro-bin centered at $z_j$, and $\bar{w}_{\text{ur}}(z_j)$ is the weighted cross-correlation of the unknown sample in photometric bin $i$ and reference sample in micro-bin $j$. As seen in the equation, we assume that within a micro-bin the galaxy bias of each sample is constant. 

It is easiest to note here that the key issue with clustering redshifts is not galaxy bias, but rather the galaxy bias evolution with redshift. In Equation \ref{nequation1}, if the galaxy biases are the same for all $z_j$ (even if unknown), they will effectively cancel out when all the $n_{\text{u},i}(z_j)$ are combined, since the total number of galaxies in the $i$-th bin is known. If the galaxy biases change with redshift within a single photometric bin though, they will not cancel out and will distort the estimated $n(z)$.

\subsection{Correcting for Galaxy Bias}
\label{sec:32}

We can get closer to solving for $n_{\text{u},i}(z_j)$ in Equation \ref{nequation1} by using auto-correlations of each sample binned by the micro-bin $j$. The weighted auto-correlations for the samples, again assuming a single galaxy bias value for the micro-bin are:

\begin{equation}
\label{autocorrelation}
\bar{w}_{\text{rr}}(z_j)=b_{\text{r}}(z_j)^2 \bar{w}_{\text{mm}}(z_j) \int n_{\text{r},j}(z)^2 dz, 
\end{equation}

\begin{equation}
\label{autocorrelation2}
\bar{w}_{\text{uu},j}(z_j)=b_{\text{u}}(z_j)^2 \bar{w}_{\text{mm}}(z_j) \int n_{\text{u},j}(z)^2 dz.
\end{equation}

We note that in Equation \ref{autocorrelation2} we have introduced new quantities, $\bar{w}_{\text{uu},j}$ and $n_{\text{u},j}$. These are quantities related to an unknown sample (i.e., a DES sample) binned in a micro-bin $j$. Our previous equations had $n_{\text{u},i}$ which relates to the unknown sample binned by photometric bin, $i$. These larger bins again correspond to the bins used by DES cosmology analyses, and thus what we ultimately want to figure out. However, as laid out in this section, we sometimes need to measure properties of this sample in smaller redshift slices (i.e., $n_{\text{u},j}$) to ultimately figure out $n_{\text{u},i}$. We also note that these $n(z)$ refer to the true (spectroscopic) redshift distribution. Thus for example, $n_{\text{u},j}(z)$ is the true redshift distribution of galaxies binned into micro-bin $j$ by \emph{photometric} redshift. So for example, $n_{\text{u},j}(z)$ in micro-bin $z_{\text{pz}} \in [0.2,0.22]$ will extend beyond $z=0.2$ and $z=0.22$ in its true redshift distribution.

If spectroscopic redshifts are obtained, in the limit of a large number of galaxies, galaxy distributions tend to be fairly flat across the small redshift range of the micro-bins ($dz=0.02$). In this limit, the normalized $n^2$ in the integrals of Equations \ref{autocorrelation}-\ref{autocorrelation2} is the same for all distributions and can be dropped. For spectroscopic (true) redshifts only, we can use Equations \ref{nequation1}-\ref{autocorrelation2} for an expression for $n_{\text{u},i}(z_j)$ in terms of measurable correlation functions:

\begin{equation}
\label{bothauto}
n_{\text{u},i}(z_j) \propto \frac{\bar{w}_{\text{ur}}(z_j)}{\sqrt{\bar{w}_{\text{rr}}(z_j) \bar{w}_{\text{uu,spec}}(z_j)}} \ \ \ (\text{w/spec-$z$ only}).
\end{equation}

However, the assumption that $n_{\text{u},j}^2$ is flat in Equation \ref{autocorrelation2} is almost certainly wrong since the unknown sample only has photometric redshifts. If the unknown sample is binned by \emph{photometric} redshift into  micro-bin $j$, the $n_{\text{u},j}^2$ in Equation \ref{autocorrelation2} will span the entire \emph{true} redshift range of that sample, which will extend beyond $dz=0.02$.

We can relate the theoretical auto-correlation of the unknown sample at $z_j$ if it could be binned by spectroscopic redshift, to the measurable auto-correlation of the unknown sample binned by $z_{\text{ph}}$ (photometric redshift) in micro-bin $j$:

\begin{equation}
\label{photozcorr}
\bar{w}_{\text{uu},\text{spec}}(z_j) \propto \bar{w}_{\text{uu,pz}}(z_j)  \frac{\int{n_{\text{spec},j}(z)^2 dz}}{\int{n_{\text{u},j}(z)^2 dz}}
\end{equation}

\noindent where $n_{\text{spec},j}$ and $\bar{w}_{\text{uu},\text{spec}}$ are respectively the true redshift distribution, and theoretical auto-correlation of the unknown sample if it could be binned by spectroscopic redshift into micro-bin $j$. Similarly, $n_{\text{u},j}$ and $\bar{w}_{\text{uu,pz}}$ are respectively the true redshift distribution and measurable auto-correlation of the unknown sample when binned by photometric redshift into the micro-bin $j$. 

This equation was used to solve for $n_{\text{u},i}(z_j)$ in determining DES Year-1 clustering redshifts in \cite{cawthon18}. Simulations were used to estimate both integrals. Again, for spectroscopic samples, $n(z)$ over a micro-bin tends to be flat and the upper integral can be dropped out. The bottom integral, the true redshift distribution of the galaxies binned in micro-bin $j$ by photo-$z$, is the main unknown. It essentially measures the photo-$z$ scatter at redshift $z_j$, with more scatter producing a wider distribution, and smaller value of $n^2$.

In a change from \cite{cawthon18}, we attempt to evaluate the photo-$z$ scatter effect in Equation \ref{photozcorr} empirically by using clustering redshift measurements on the photometric galaxies binned in each micro-bin $j$ by photometric redshift, as an estimate of $n_{\text{u},j}(z)$. We will assume in this narrower redshift range spanned by $n_{\text{u},j}(z)$, that we can approximate $b_{\text{u}} \bar{w}_{\text{mm}}$ as constant. From that approximation and Equations \ref{nequation1}-\ref{autocorrelation}, we have:

\begin{equation}
\label{nanobins}
n_{\text{u},j}(z) \propto \frac{\bar{w}_{\text{ur}}(z)}{\sqrt{\bar{w}_{\text{rr}}(z)}} \ \ \ (\text{assume} \ b_{\text{u}} \bar{w}_{\text{mm}}=\text{const.}).
\end{equation}

Measurements of $n_{\text{u},j}(z)$ are  noisier than evaluating $n_{\text{u},i}(z)$ (Equation \ref{bothauto}). We are dividing up the DES photometric sample into the smaller micro-bins $j$ than the main photometric bins, $i$. Furthermore, since $n_{\text{u},j}(z)$ is narrower than $n_{\text{u},i}(z)$, we evaluate Equation \ref{nanobins} on even smaller bins than the $dz=0.02$ micro-bins. These `nano-bins' are either $dz=0.01$ \text{or} $0.005$ depending on the signal to noise. However, these further subdivisions make each measurement noisier. In order to reduce the computations and not propagate as many noisy data points, we make the approximation that the integral of $n^2(z)$ can be estimated by simply the inverse of the standard deviation of $n(z)$:

\begin{equation}
\label{stnddev}
\int{n_{\text{u},j}(z)^2} dz \approx \frac{1}{\sigma_j}
\end{equation}

\noindent where $\sigma_j$ is the standard deviation of the redshift distribution. 

We note that photometric redshift scatter is often approximated as a Gaussian function (\citealt{cawthonthesis}, \citealt{lsstbook}). If $n(z)$ is a Gaussian (i.e., $n(z) \approx \frac{1}{\sigma \sqrt{2 \pi}} e^{-\frac{1}{2}(\frac{z-\mu}{\sigma})^2}$) with $\mu \ \text{and} \ \sigma$ the mean and standard deviation respectively) the integral of $n^2$ directly evaluates to $\propto 1/\sigma$. Thus the approximation of Equation \ref{stnddev}, particularly in trying to estimate photo-$z$ scatter, seems appropriate. \footnote{Even in the case of a perfect Gaussian redshift distribution, Equation \ref{stnddev} would not be exact due to the finite width of the micro-bins.} We also note that if there is linear galaxy bias evolution across even the `micro-bin' measurement (i.e., $b_{\text{u}} \bar{w}_{\text{mm}}$ varies across the small redshift range, just as we expect it to for the larger, main photometric bins), this would have far less impact on the standard deviation, $\sigma$, than it will on the mean redshift.

\begin{figure}
\begin{center}
\includegraphics[width=0.5 \textwidth]{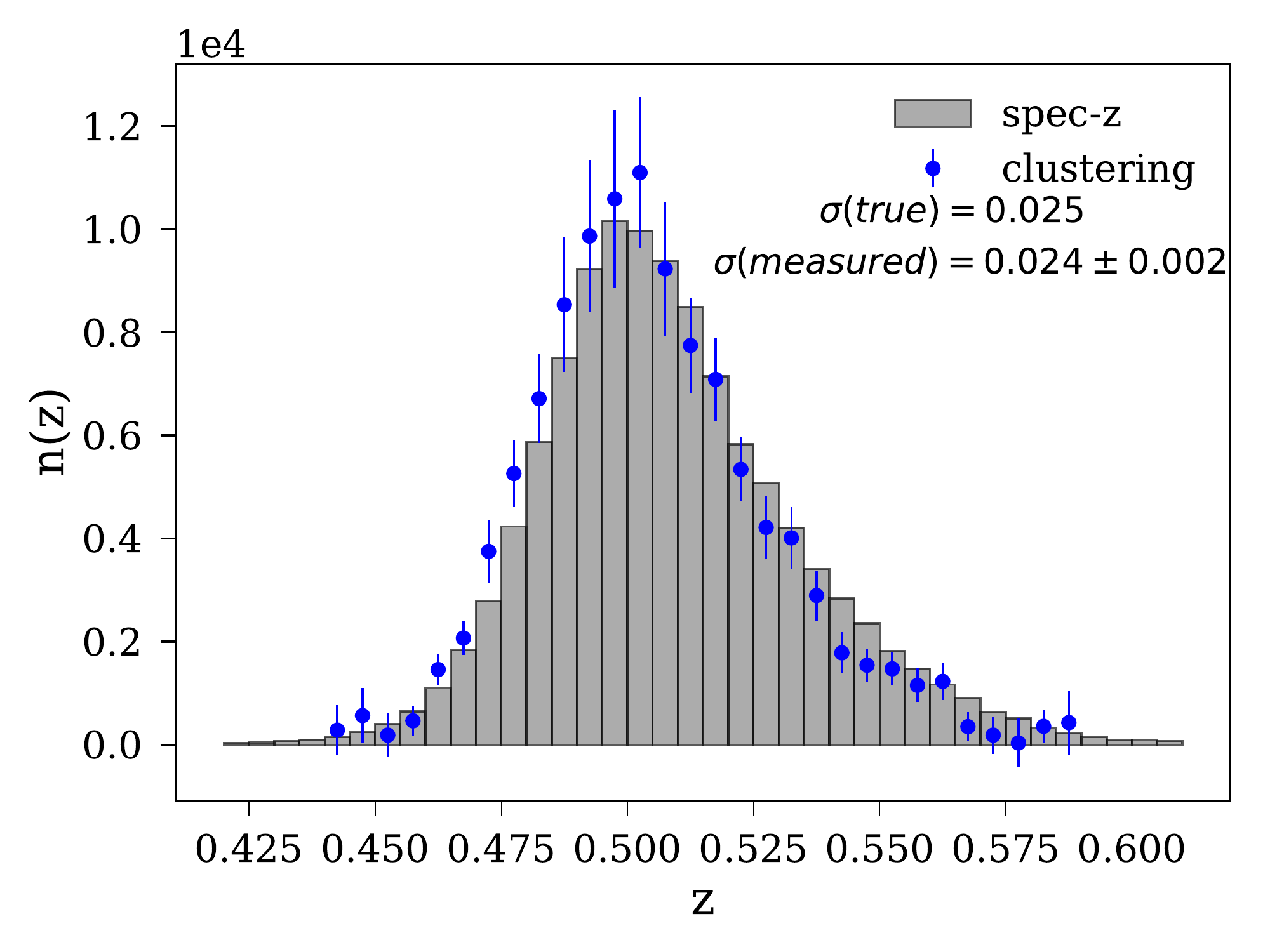}
\end{center}
\caption{An example of clustering redshift measurements on a `micro-bin' of size $dz=0.02$ in a simulation. Shown are the true redshift distribution, and the one measured from clustering redshifts, on simulated \redmagic galaxies with peak photo-$z$ probability between $z=0.49-0.51$. The goal of this particular measurement is to measure the standard deviation, $\sigma$, of the $n(z)$. The standard deviation serves as a proxy for estimating how much an auto-correlation of this same micro-bin of \redmagic will be reduced compared to the spectroscopic case (where the entire $n(z)$ would be entirely between 0.49 and 0.51). To do this measurement, the spectroscopic sample (DMASS) is divided up into smaller `nano-bins' of size $dz=0.005$ in order to get more data points for the measurement.}
\label{fig:nanobinmeasure}
\end{figure}

\begin{figure}
\begin{center}
\includegraphics[width=0.5 \textwidth]{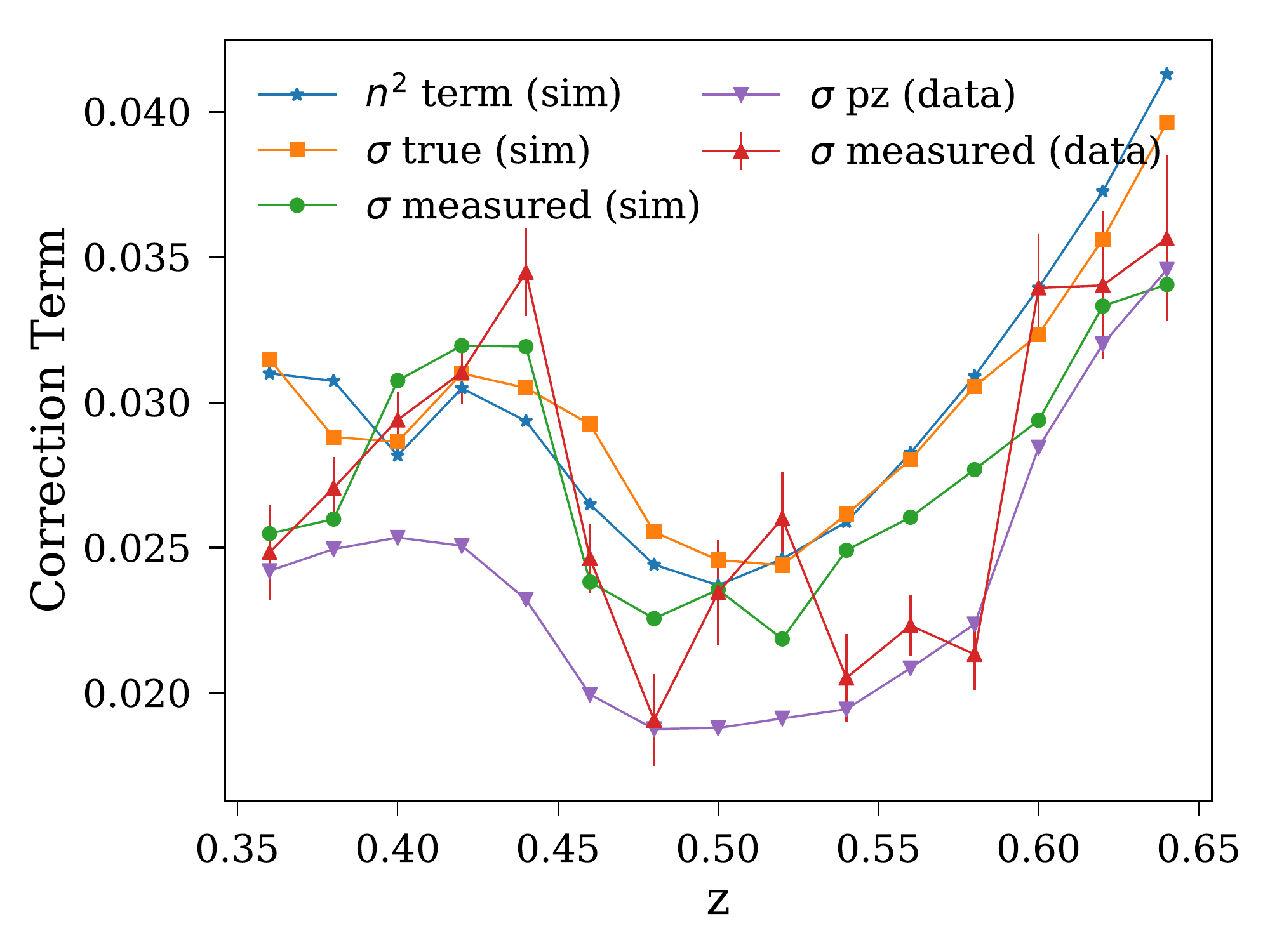}
\end{center}
\caption{Various approximations of the `correction term' needed to estimate the auto-correlation of a sample micro-binned by true (spectroscopic) redshift from the auto-correlation of the sample micro-binned by photometric redshift. The correction term is either $\sigma$, the standard deviation of an $n(z)$ estimate, or the `$n^2$ term', $1/\int n_{\text{u},j}^2$ from Equation \ref{photozcorr}. The `$n^2$ term' curve is normalized to the `$\sigma$ (true)' data points. Specifically, these $\sigma$ estimates represent the standard deviation of the true redshift distribution when binned by photo-$z$. This value is either calculated exactly (simulations), measured by clustering redshifts on smaller `nano-bins' (on either simulations or data) or calculated from the photometric redshift estimates (data). We quantify the agreement of some of these approaches in Section \ref{sec:simtests}. The $\sigma$ pz curve is the $\sigma$ estimated from the photo-$z$ algorithm itself (i.e., from $n^{\text{pz}}(z)$)}.The sample shown is the high density \redmagic samples from $z_{\text{ph}}=0.35-0.65$ in both simulations and data.
\label{fig:sigmacomparisons}
\end{figure}

Using simulations, we validate the efficacy of the approximation in Equation \ref{stnddev}. In Figure \ref{fig:nanobinmeasure}, we show an example of measuring $\sigma_j$, the standard deviation in one of the micro-bins, by doing clustering redshift measurements in the smaller `nano-bins'. In Figure \ref{fig:sigmacomparisons}, we show comparisons of the $\sigma_j$ and $1/\int n^2$ correction terms measured in different ways in simulations and in data. We quantify the differences of each approximation (Equations \ref{bothauto}-\ref{stnddev}) in our tests in Section \ref{sec:simtests}.

Using Equations \ref{bothauto}-\ref{stnddev}, we get our estimate for $n_{\text{u}}(z_j)$:

\begin{equation}
\label{bothautomod}
n_{\text{u}}(z_j) \propto \frac{\bar{w}_{\text{ur}}(z_j)}{\sqrt{\bar{w}_{\text{rr}}(z_j) \bar{w}_{\text{uu,pz}}}(z_j) \sigma_j} .
\end{equation}

\noindent Since $\sqrt{\bar{w}_{\text{uu,pz}}(z_j)\sigma_j}$ is a noisy approximation of $\sqrt{\bar{w}_{\text{uu,spec}}(z_j)}$, we approximate it with a power law, as was done in \cite{davis18} and \cite{cawthon18}:

\begin{equation}
\label{powerlaw}
\sqrt{\bar{w}_{\text{uu,spec}}(z_j)} \approx \sqrt{\bar{w}_{\text{uu,pz}}(z_j) \sigma_j} \propto (1+z)^\gamma .
\end{equation}

We test the accuracy of our methodology in various steps in Section \ref{sec:simtests}. Specifically, we test the method when using spectroscopic redshifts (Equation \ref{bothauto}), the power law approximation  when using spectroscopic redshifts (Equation \ref{powerlaw}), and the approximate solutions when using photometric redshifts (Equations \ref{photozcorr} and \ref{bothautomod}).

\subsection{Estimating Photometric Redshift Bias (1-Parameter Fit)}
\label{sec:33}
Thus far, this work has focused on the clustering redshift measurements themselves and their veracity. We now briefly discuss how specifically these measurements are used to calibrate the photometric redshift distributions used in \cite{y3-3x2ptkp} and related papers.

Our general strategy is to use a calibrated photo-$z$ distribution rather than clustering redshifts directly. This strategy is formed from a belief that the clustering redshifts are more accurate overall than the photometric estimate, but are not reliable in the tails due to noise and magnification effects. The clustering redshifts can also give unphysical, negative $n(z)$ measurements in the tails. A calibrated photo-$z$ distribution, $n^{\text{pz}}(z)$ (where we now use an upper index to distinguish from clustering or spectroscopic estimates of $n(z)$) prediction is thus preferable to using clustering redshifts directly.

For our fiducial plan, we assume a single shift parameter, $\Delta z$, is enough to calibrate the photo-$z$ distribution. This parameter is essentially the photo-$z$ bias. Our final clustering distribution, $n_{\text{u}}(z)$, comes from Equations \ref{bothautomod} and \ref{powerlaw},  estimated for each $i$-th photometric bin, with data points in each micro-bin, $j$. Our final step is thus to find the shifted $n^{\text{pz}}(z)$ that matches the mean of our $n_{\text{u}}(z)$ from clustering redshifts. Specifically, we find the shift, $\Delta z$ that satisfies:

\begin{equation}
\label{meanfitting}
\frac{\int_{z_{\text{min}}}^{z_{\text{max}}} z \ n^{\text{pz}}(z-\Delta z) \ dz}{\int_{z_{\text{min}}}^{z_{\text{max}}} n^{\text{pz}}(z-\Delta z) \ dz}  =   \frac{\int_{z_{\text{min}}}^{z_{\text{max}}} z \ n_{\text{u}}(z) \ dz}{\int_{z_{\text{min}}}^{z_{\text{max}}} n_{\text{u}}(z) \ dz} .
\end{equation}

As in \cite{cawthon18} and \cite{gatti18}, we set $z_{\text{min}}$ and $z_{\text{max}}$ to be at 2.5$\sigma$ from the peak of the clustering redshift distribution. This cuts the tails of $n_{\text{u}}(z)$ from being used in the calculation for the correct shift. The tails of a clustering redshift estimate can be noisy, with negative signals being difficult to calibrate. The tails can also be affected by magnification as we discuss in more detail in Section \ref{sec:magnification}.

\subsection{Estimating Photometric Redshift Bias+Stretch (2-Parameter Fit)}
\label{sec:44}
In this work, we will also use a 2-parameter fit to calibrate the photo-$z$ distribution with the clustering redshift measurements. In general, this fit will work in cases where a 1-parameter shift of the photo-$z$ distribution is a poor fit to the clustering data, i.e., the shapes of the distributions disagree.

For this fit, we use the $\Delta z$ shift parameter from the 1-parameter fit, as well as a `stretch' parameter, $s$. The stretch parameter is included by shifting the photo-$z$ distribution such that its mean is centered at $z=0$, then re-scaling the z axis by a factor $s$, and finally shifting back to $z_{\rm mean}$. The functional form of this is given by 

\begin{equation}
\label{stretch}
    n_{\rm 2-param}(z) = \frac{1}{s} n_{\rm pz}\left(\frac{z-z_{\rm mean}-\Delta z}{s} + z_{\rm mean}\right),
\end{equation}

\noindent where $s$ is the new stretch parameter, equal to $1$ if the width of the photo-$z$ and clustering-$z$ are the same, and $\Delta z$ is the usual shift parameter. We refer to this as the 2-parameter model. We  apply a $\chi^{2}$ least squares fitting of $s$ and $\Delta z$ to the clustering redshift results. To account for the galaxy bias correction in a manner similar to our fiducial methods of Section \ref{sec:methods}, we propagate $\gamma$ (Equation \ref{powerlaw}) into the clustering redshift $n(z)$ and covariance when doing the $\chi^{2}$ fit. In Section \ref{sec:5chi}, we test in simulation the $\chi^{2}$ fit method and two other methods of fitting for $\Delta z$ and $s$.

\section{Testing Methodology with Simulations}
\label{sec:simtests}

We validate various steps in our methodology, and estimate any associated biases or systematic uncertainties with those steps, with tests in simulations. As described in Section \ref{sec:datasets}, we use the Buzzard simulations, and simulated samples of the DES \redmagic and \maglim galaxies, BOSS CMASS galaxies, and eBOSS ELG galaxies. We divide the simulated DES galaxies into 6 samples, corresponding to bins 2, 3, and 4 for \redmagic, and bins 2, 3, and 4 for the \maglim sample. These samples were chosen for their redshift overlap with the two simulated BOSS/eBOSS samples.

We evaluate the different steps by testing our methodology on each of the six samples. From the six results, we then fit for bias and systematic uncertainty (on top of statistical uncertainties for the correlation functions). We describe the evaluation step in more detail at the end of this Section.

\subsection{Test 1: Testing methodology with spectroscopic redshifts}
\label{sec:51}

Our first test evaluates the accuracy of Equation \ref{bothauto}, the solution for $n(z)$ when using all spectroscopic measurements. In this scenario, auto-correlations of both the unknown and reference samples can be used to calibrate the impact of galaxy bias evolution with redshift across a photo-$z$ bin. Since the DES samples are photometric, we can only carry out this measurement on the simulations. This test is still useful to isolate performance of the method, including galaxy bias corrections from auto-correlations, before evaluating any of the effects associated with photometric redshifts.

\subsection{Test 2: Approximating galaxy bias correction with a power law}
\label{sec:52}
In Equation \ref{powerlaw}, we approximate $\bar{w}_{\text{uu}}(z)$ (or proxies of it) as a power law: $(1+z)^\gamma$. In Test 2, we test any biases in fitting the auto-correlation to a power law. To isolate the effects of this approximation from the effects of photometric redshifts, we do this test on the auto-correlation of the simulated DES samples, when using \emph{spectroscopic} redshifts. We note that in the following three tests (Tests 3,4,5), the proxies for the auto-correlation, $\bar{w}_{\text{uu}}(z)$ are approximated as a power law as well.

\subsection{Test 3: Galaxy bias correction using photometric redshifts and redshift scatter model}
\label{sec:53}

In Test 3, we test how well Equation \ref{photozcorr} corrects for effects of photo-$z$ scatter in the simulations. The integrals in Equation \ref{photozcorr}, which describe how the redshift distribution changes when binned by photometric or spectroscopic redshift, can only be evaluated in simulations with true redshift information. In the Year 1 DES results, \cite{cawthon18} used this calculation from simulations for the final correction to the clustering redshift results. We also fit Equation \ref{photozcorr} to a power law for this test. 

\subsection{Test 4: Galaxy bias correction using standard deviation}
\label{sec:54}

In Test 4, we test the approximation of Equations \ref{stnddev}-\ref{bothautomod}, using the standard deviation of a redshift distribution, $\sigma_j$, as an approximation in the photo-$z$ correction of Equation \ref{photozcorr}. In this test, we calculate $\sigma_j$ for the DES samples exactly from their true redshifts in the simulation. Our approximation of $\bar{w}_{\text{uu,spec}}(z_j)$ in this test is then $\bar{w}_{\text{uu,pz}}(z_j) \sigma_j$. We again fit these new estimates of $\bar{w}_{\text{uu,spec}}(z_j)$, for each micro-bin $j$ within the photometric bin, $i$, to a power law.

\subsection{Test 5: Galaxy bias correction using standard deviation inferred from clustering}
\label{sec:55}
In Test 5, we test the last step in calculating a proxy for $\bar{w}_{\text{uu}}(z_j)$. We again calculate $\sigma_j$, but this time from estimating the redshift distribution of the photo-$z$ microbinned DES sample by using cross-correlations with BOSS/eBOSS samples on even smaller `nano-bins' (Figure \ref{fig:nanobinmeasure}). We estimate $\sigma_j$ from these results and again test Equations \ref{stnddev}-\ref{bothautomod}. Unlike Tests 3 and 4, Test 5 describes a measurement that can be done on the data, without true redshift information.

\begin{figure*}
\begin{center}
\includegraphics[width=1.0 \textwidth]{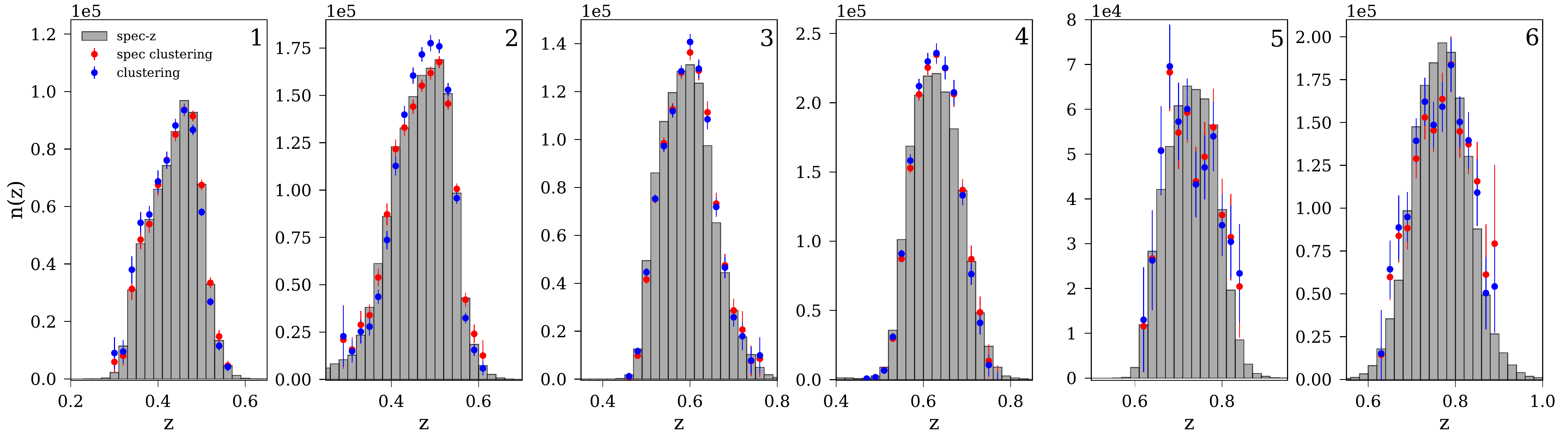}
\end{center}
\caption{Clustering redshift estimates on simulated DES galaxy samples. The first, third and fifth samples are simulated \redmagic redshift bins, and the others are simulated \maglim samples. The first four bins are cross-correlated with the simulated DMASS sample. The last two are correlated with the simulated ELG sample. Shown are the true (spec-$z$) redshift distributions, the clustering measurements that could be derived if both samples had spectroscopic redshifts (`spec clustering, Test 1'), and the clustering measurements described in `Test 5', which is also the procedure done on the data. The simulated eBOSS sample runs out of galaxies around $z=0.9$, limiting the higher-$z$ bins shown here.}
\label{fig:simwz}
\end{figure*}

\begin{figure}
\begin{center}
\includegraphics[width=0.5 \textwidth]{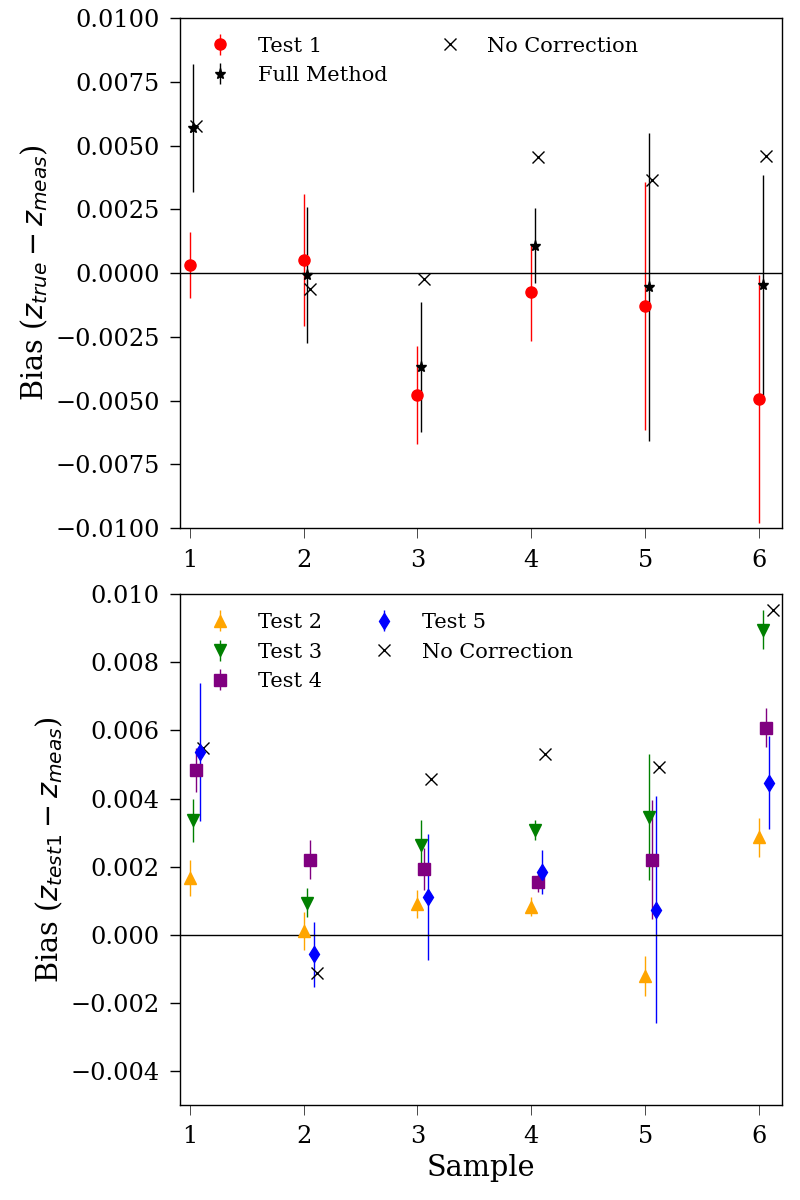}
\end{center}
\caption{Results of testing different steps of the methodology on six different simulated samples. The tests are described throughout Section \ref{sec:simtests}. Top: Tests comparing the measured $z_{\text{mean}}$ with the true one. Test 1 compares the method if spectroscopic redshifts were available for the photometric sample auto-correlations. Full method is the method we can do on the data.  Bottom: Tests 2-5 focus on different correction steps or techniques for simulating the true (spectroscopic) photometric sample auto-correlation. The bias shown is with respect to the mean redshift from Test 1 in order to isolate bias due to the correction alone. Test 5 represents the full method that can be done on data. For each panel, the bias with `no corrections' shows the results if no attempt is made to use or measure the photometric sample auto-correlation. The six test samples are in order of redshift. In order, they are simulated versions of (1) \redmagic Bin 2,  (2) \maglim Bin 2, (3) \redmagic Bin 3, (4) \maglim Bin 3, (5) \redmagic Bin 4, (6) \maglim Bin 4. The simulated samples are described in Section \ref{sec:sims}.}
\label{fig:wztests}
\end{figure}

\subsection{Summary of tests on photo-$z$ bias}
\label{sec:47}
For each test in the simulations, we compute the clustering redshift measurements. We show in Figure \ref{fig:simwz} for each of the six simulated samples: the true redshift distribution, the clustering estimate in Test 1 (using auto correlations of DES samples using true redshifts), and the clustering estimate in Test 5, a procedure for calibrating the galaxy bias effects that can be done on the data.

We calculate the mean redshift in each case for the clustering redshifts. We summarize the accuracy of each test based on the measured mean redshift in Figure \ref{fig:wztests}. For each of our six simulated DES test samples, we plot the `bias' of each step individually, each represented by a different test. For Test 1, the method if spectroscopic redshifts are available for the unknown sample, the `bias' is the true mean redshift of the photometric bin (calculated in the range where there are clustering estimates) minus the inferred mean redshift. For Tests 2-5, we plot as `bias' the inferred mean redshift of Test 1 minus the inferred mean redshift of the test in question. We do this since tests 2-5 all involve replacing $w_{\text{uu}}$ in some way. Comparing them to Test 1 thus isolates the bias of the specific approximations of $w_\text{uu,spec}$ compared to the case where $w_\text{uu,spec}$ is actually available.

The method when spectroscopic redshifts are available (Test 1, `Method') is shown to be generally accurate. Five of the six samples estimated mean redshifts which are consistent with the true mean redshift (i.e., zero bias) within the statistical error bar. The power law approximation (Test 2) also shows little bias. Tests 3,4,5, which all evaluate methods of modifying $w_\text{uu,pz}$ to estimate $w_\text{uu,spec}$, show a relatively small positive bias in all six samples, suggesting a true bias of this approximation. A positive bias means the inferred mean redshifts are too low. We also show the case of no correction, where no estimate of $w_{\text{uu}}$ is used. We see that in each case, the attempted correction of Test 3, 4 or 5, does reduce some of the bias.  

We now quantify these tests to add any necessary biases or systematic uncertainties of the method to our results later. A non-zero systematic uncertainty indicates that the overall uncertainty of the method is larger than the measured statistical uncertainty from the jackknife resamplings. For each test, we model its bias, $b$, and systematic uncertainty, $\omega$ with a two-parameter fit to the six data points in Figure \ref{fig:wztests} and do a chi-squared test to find the best fit parameters. Specifically, we calculate for each test:

\begin{equation}
\label{chisquare}
\chi^2_{\text{red}}=\sum_{x=1}^{6}\frac{(d_{x}-b)}{\sqrt{\sigma_{x}^2+\omega^2}} /(\text{dof}=4)
\end{equation}

\noindent where $\chi^2_{\text{red}}$ is the reduced chi-squared, $x$ iterates over the six test samples, $d_{x}$ is the measured bias on each sample for the given test, $\sigma_{x}$ is the measured statistical error on those biases, and dof signifies our 4 degrees of freedom (six samples - two fit parameters). For a reduced chi-squared, a value of 1 represents a good fit, a value of less than 1 indicates too good a fit (errors are overestimated) and a value of greater than 1 indicates a poor fit (errors are underestimated).

We start by evaluating a large set of $b$ parameters with $\omega=0$. If any of the resulting reduced chi-squares are less than 1, this would indicate a good fit to a bias of the step, with no uncertainty needed to be added on top of the statistical uncertainties. For each of the five tests, no $b$ values result in $\chi^2_{\text{red}}<1$ when $\omega=0$. This indicates that each step tested has some systematic uncertainty beyond the statistical uncertainties from the auto- and cross-correlation measurements. We then incrementally continue to calculate $\chi^2_{\text{red}}$ for a range of $\omega$ and $b$ values. We choose as our best fit the smallest value for $\omega$, and the corresponding $b$ that result in $\chi^2_{\text{red}}<1$. Our results are shown in Table \ref{table:test_results}. As suggested by Figure \ref{fig:wztests}, each of the biases are relatively small ($|b| \leq 0.0037$).

We choose to only use Tests 1 and 5 to estimate the bias and uncertainty we should add to our measurements of the data. Test 1 (`Method') essentially estimates biases across the full method presented in his work, modulo the complication of not having spectroscopic redshifts to estimate $w_\text{uu,spec}$. Tests 3,4,5 all measure the step of trying to estimate $w_\text{uu,spec}$. Of these three, Test 5, which measures $\sigma_j$, the width of the true redshift distribution of photometric galaxies in micro-bin $j$ from clustering redshift measurements on `nano-bins', is the only method that can be used solely from data. It happens to be that this method in the simulations (Test 5) is slightly more accurate than the other methods (Test 3 and 4) which estimate the photometric correction either from an exact calculation of $\int n(z)^2$ or estimating $\sigma_j$ exactly. Since tests 3,4, and 5 all approximate $w_\text{uu,spec}$, adding all of their uncertainties would likely be redundant and overestimate our errors. Tests 3,4 and 5 also involve a fit to a power law, so the small errors found in Test 2 are likely incorporated into the results of Test 5. Thus, deriving systematic biases and uncertainties from Tests 1 and 5 only incorporates each element of the measurements a single time.

\begin{table}
\small\addtolength{\tabcolsep}{-5pt}
\begin{center}
    \begin{tabular}{|c|c|c|c|}
      \hline
      \multicolumn{4}{|c|}{Methodology Test Results} \\
      \hline
      Name of Test & Bias & Uncertainty & In Error Budget? \\
      \hline
      Test 1: Method w/Spec-$z$ & -0.0014 & 0.0013 & Yes  \\   
      \hline
      Test 2: Power-law approx. & 0.0009 & 0.0015 & No  \\   
      \hline
      Test 3: Exact $n^2$ correction & 0.0037 & 0.0030 & No  \\   
      \hline
      Test 4: Exact $\sigma_j$ correction & 0.0032 & 0.0020 & No  \\   
      \hline
      Test 5: Clustering-$z$ $\sigma_j$ correction & 0.0021 & 0.0021 & Yes  \\   
      \hline
      Combined Errors & 0.007 & 0.0025 & -  \\  
      \hline
      Full Method Check & 0.007 & 0.0023 & -  \\ 
      \hline
    \end{tabular}
  \caption{Analysis of the tests shown in Figure \ref{fig:wztests}, by fitting parameters in Equation \ref{chisquare}. The tests are described throughout Section \ref{sec:simtests}. We note the uncertainties represent uncertainty to be added to a step in the method. For example, Test 5 has more statistical uncertainty already in its step than Test 4, so the results do not imply Test 4 has more total uncertainty. As described in the text, we choose to incorporate the biases and uncertainties of Tests 1 and 5 to not double count uncertainties in any step of our methodology of solving for the redshift distribution, $n_{\text{u}}(z)$. Combined errors adds the counted biases and uncertainties (the latter in quadrature). The full method check is the bias and uncertainty found when not breaking up the analysis in different steps.}
  \label{table:test_results}
\end{center}
\end{table}

From Table \ref{table:test_results}, we take the results from Test 1 and Test 5. We add the biases linearly and the systematic uncertainties in quadrature. This results in adding a bias of +0.0007 and a systematic uncertainty of 0.0025 to each of our measurements. We will see in Section \ref{sec:results} that these systematic errors are generally similar to or smaller than our inferred biases and statistical uncertainties on the mean redshift (or equivalently, the $\Delta z$ parameter). 

We also show in Table \ref{table:test_results} a calculation of a `Full Method Check'. This calculation is simply taking our fiducial estimate of clustering redshifts (used in Test 5) and comparing to the true mean redshift, rather than to the results of Test 1, which is what Test 5 does. In this case of the full method check, the intention is not to estimate a bias of a single step, but to estimate the bias of the entire method, end to end. We see in the table, that the resulting bias and uncertainty are very similar to `added' bias and uncertainty values of +0.0007 and 0.0025. We can thus conclude that whether we broke up the method into different steps or not would not have notably impacted our derived bias and uncertainty.

In Appendix \ref{sec:testsappendix}, we briefly discuss a few alternative ways of evaluating these tests, as well as the assumption of the six samples each being independent tests of the method. We find very minor differences in bias and uncertainty in all cases explored.

\subsection{Tests of 2-Parameter Fits}
\label{sec:5chi}
In this section, we test the accuracy of the 2-parameter fit (Equation \ref{stretch}). The second parameter, the stretch, changes the width of the distribution. The galaxy bias correction, at least in the power law form (Equation \ref{powerlaw}) has very little impact on the stretch parameter, so our previous approach of breaking up the method into various steps is less well motivated. Therefore, we estimate the accuracy of the 2-parameter fit using just the `full method' estimate, where the final $\Delta z$ and $s$ parameters from our fiducial clustering redshift estimate (Equations \ref{bothautomod} and \ref{powerlaw}) are compared to the true values. As seen in Table \ref{table:test_results}, this test produced nearly identical results for the 1-parameter fit as when we evaluated different steps separately.

We test a few different ways of fitting the 2-parameter model, each using the $\Delta z$ and $s$ parameters from Equation \ref{stretch}. The first, as mentioned in Section \ref{sec:44}, is a $\chi^2$ fit to the clustering redshift data points, selecting $\Delta z$ and $s$ to change the photo-$z$ distribution such as to be as close to the clustering redshifts as possible. The second method is a more natural extension of the 1-parameter fit. It selects $\Delta z$ in the same way as the 1-parameter fit. Then, it selects the $s$ parameter that makes the photo-$z$ distribution have the same standard deviation, $\sigma$, as the clustering data. We call it the `shift then stretch' (STS) method. This method fits to parameters, rather than fit to all the points, like the 1-parameter fit. It will also give the same $\Delta z$ constraints as the 1-parameter fit, while the $\chi^2$ method may not. We also try a third method, which we call `Mix'. It first computes the 1-parameter shift (so also gives same results on $\Delta z$ as STS). Then, with a fixed $\Delta z$, it does a $\chi^2$ fit for the stretch.

\begin{figure}
\begin{center}
\includegraphics[width=0.5 \textwidth]{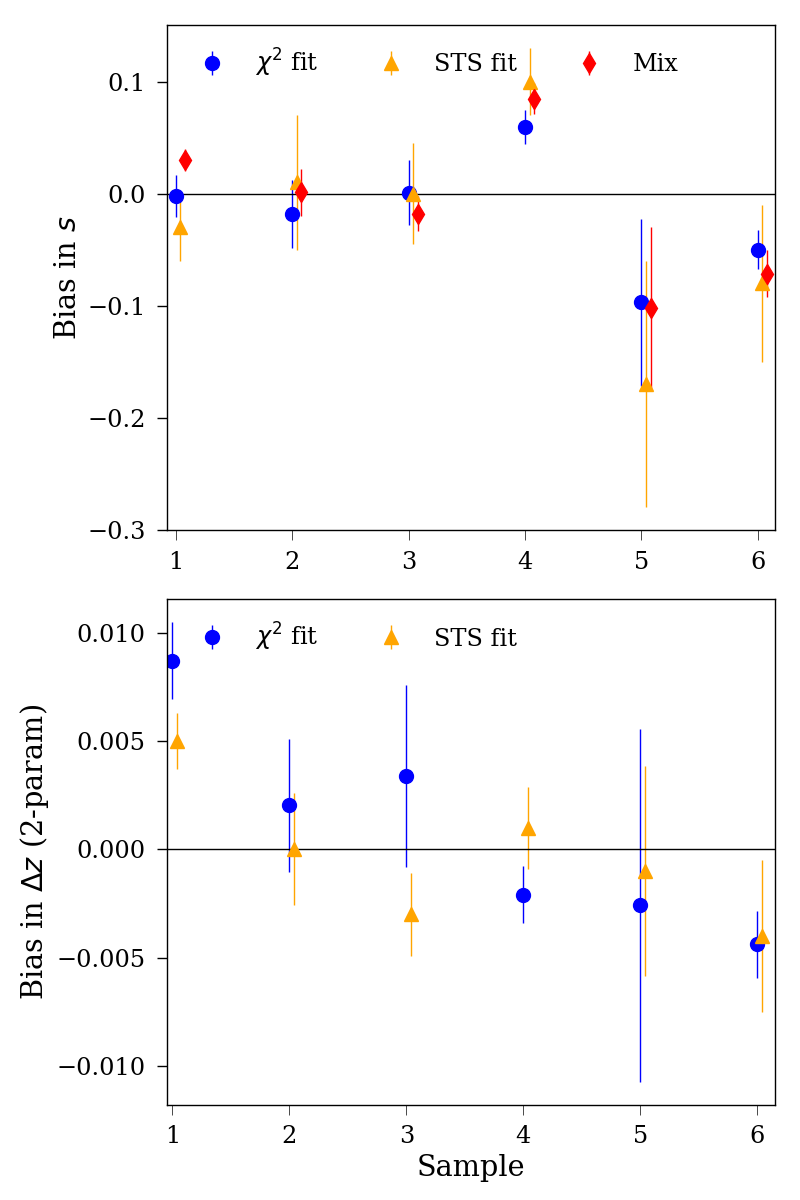}
\end{center}
\caption{Results of testing the three different 2-parameter fit methods, $\chi^2$, STS (shift then stretch), and Mix (shift, then $\chi^2$ fit for stretch) on the six simulated samples. We show the biases in measured $\Delta z$ and $s$ parameters compared to the true values that make the photo-$z$ distribution closest to truth. As seen, the STS method has smaller errors and is more accurate in getting $\Delta z$ correct, but has larger errors and is more biased on the stretch parameter, $s$. The $\chi^2$ method is the most accurate and has the smallest errors on the stretch parameter. Mix gives similar results to $\chi^2$ for the stretches, but is still a bit more inaccurate for that parameter. Mix has the same $\Delta z$ estimate as STS. We fit for a `method uncertainty' for each method, on each parameter, based on these results in Table \ref{table:test_results_2param}.}
\label{fig:wztests2param}
\end{figure}

To test the three methods of fitting, we compute $\Delta z$ and $s$ for each of the six simulated samples, given their simulated photo-$z$ distributions. We also calculate the true $\Delta z$ and $s$ parameters that would make the photo-$z$ distribution most closely match the true distribution. Given these estimated and true $\Delta z$ and $s$ parameters, we compute biases on each parameter, for each method on each sample. The results are shown in Figure \ref{fig:wztests2param}.

As we did for the 1-parameter fit (photo-$z$ bias only), we evaluate a `method error' by comparing the estimated parameters with their true values. We again use Equation \ref{chisquare}, though we decide to use a more agnostic model of no bias, but just an added uncertainty ($\omega$ in Equation \ref{chisquare}). We chose this since in Figures \ref{fig:simwz} and \ref{fig:wztests2param}, there is some evidence of a directional bias to the stretch parameter based on the signal to noise, with stretches (widths) being overestimated with noisy data (samples 5 and 6). In the first four bins, with better signal, there is a small preference for an opposite bias on the stretch, with a preference for narrower distributions than truth. Without more simulated samples to investigate these relationships, a no bias fit seemed most conservative, likely resulting in larger uncertainty than otherwise. 

\begin{table}
\small\addtolength{\tabcolsep}{-5pt}
\begin{center}
    \begin{tabular}{|c|c|c|c|}
      \hline
      \multicolumn{3}{|c|}{2-Parameter Methodology Test Results} \\
      \hline
      Method & Parameter & Uncertainty \\
      \hline
      $\chi^2$ & $\Delta z$ & 0.0044 \\   
      \hline
      $\chi^2$ & $s$ & 0.038 \\   
      \hline
      STS & $\Delta z$ & 0.0025 \\
      \hline
      STS & $s$ & 0.060 \\   
      \hline
      Mix & $s$ & 0.052 \\
      \hline
    \end{tabular}
  \caption{Analysis of the results shown in Figure \ref{fig:wztests2param}. We compare the estimates of $\Delta z$ and $s$ with the true values in the simulation with Equation \ref{chisquare} (setting b=0 and just solving for an uncertainty, $\omega$.)}
  \label{table:test_results_2param}
\end{center}
\end{table}

In Table \ref{table:test_results_2param}, we show the estimated `method uncertainty' of each parameter for each 2-parameter method. These uncertainties are to be added to the statistical errors of a given method. As can also be inferred from Figure \ref{fig:wztests2param}, the STS (shift then stretch) method does better at getting the mean redshift, but is significantly less accurate than $\chi^2$ in recovering the stretch parameter. The $\chi^2$ method is the most accurate and has the smallest errors on the stretch parameter. The Mix method gives similar results to $\chi^2$, but is still notably less accurate for the stretch parameter. It is more accurate than the STS method for the stretch though.

We note that for sample 6 (simulated \maglim bin 4), for only the 2-parameter tests in Figure \ref{fig:wztests2param}, we used larger clustering scales, 0.5-4 Mpc, as we do for the noisier \maglim bins in the data. Without this, the sample 6 stretch is significantly more negatively biased for each fit, and drives the method uncertainty for $\chi^2$ to 0.07, significantly deviant from the uncertainty of 0.035 when fitting the first five bins. This change in scales would impact all of the 1-parameter results by less than 0.001.

In principle, any of these methods with the extra uncertainties from these tests should be unbiased. When we look at the overall error budget the three methods give on measurements of data, we find the STS and Mix methods have slightly smaller uncertainty on $\Delta z$, but notably larger uncertainty on the stretch (particularly STS). Based on these results, we proceed with the $\chi^2$ method as our fiducial 2-parameter fit method. We reiterate that each method is just a different way of fitting to the clustering redshift data, so the similar biases in Figure \ref{fig:wztests2param} are expected. We note that we also tried fitting for the parameters in Tables \ref{table:test_results}-\ref{table:test_results_2param} using a maximum likelihood formalism and found consistent results in each case.

In Appendix \ref{sec:densitydependenceappendix}, we describe tests to analyze possible density dependence of the systematic uncertainties derived from this Section. We do find some indication of a density dependence which would result in the uncertainties presented here being underestimated for the data being used. We discuss these results and implications in that Appendix, finding even in a pessimistic case, the impact on cosmological results is minimal.

\section{Results}
\label{sec:results}

We show our clustering redshift estimates for the 5 \redmagic and 6 \maglim redshift bins in Figures \ref{fig:redmagicresults}-\ref{fig:maglimresults}. In those figures, we also show the predicted redshift distributions from photometric redshift algorithms. For \redmagic, this is provided by the \redmagic algorithm itself. For \maglim, this is provided by the \dnff photo-$z$ algorithm, using its full probability distribution function (PDF). In each figure, the solid blue points indicate the $2.5 \sigma$ range of the clustering signal which we use to calculate the mean redshift from clustering. The faded gray points are shown, but not used. For each bin, we compute the 1-parameter fit, where we find $\Delta z$, a shift parameter to make the photometric and clustering distributions match in mean redshift (Equation \ref{meanfitting}). We also compute the 2-parameter fit, where a $\chi^2$ fit to the data simultaneously fits for a $\Delta z$ and a stretch parameter, $s$. This 2-parameter fit is also done on the $2.5 \sigma$ range of the clustering signal. 

The best 1- and 2-parameter fits and uncertainties are listed in Tables \ref{table:redmagic}-\ref{table:maglim_2param}. Also shown is the $\chi^2$ value between the fit and the clustering redshift data points. The listed uncertainties in the fits, as well as the error bars in the figures, include contributions from statistical and systematic uncertainties. The statistical uncertainties come from the cross-correlation of unknown and reference samples ($\bar{w}_{\text{ur}}$) and auto-correlations of the reference sample ($\bar{w}_{\text{rr}}$) in Equation \ref{bothautomod}. For the 1-parameter fits, the systematic uncertainty is calculated on the derived mean redshift specifically and has two sources. The `method uncertainty' of 0.0025 derived from Section \ref{sec:simtests}, and uncertainty in the calculations of the auto-correlations of the unknown (DES) sample and the calculation of the standard deviation parameter ($\bar{w}_{\text{uu,pz}} \sigma_j$). This uncertainty is propagated into an uncertainty on $\gamma$ in fitting that quantity to a power law (Equation \ref{powerlaw}). The auto-correlations and the power law fits are shown in Figure \ref{fig:desauto}. These two sources of systematic uncertainty are added in quadrature. The total uncertainty comes from adding the systematic and statistical uncertainty in quadrature. 

For the 2-parameter fit, the uncertainty from the power law is propagated into the data points before the fit is done, thus the statistical error incorporates the power law fit. The only remaining systematic uncertainty is the `method' uncertainty from Table \ref{table:test_results_2param} which is added in quadrature to each parameter's statistical uncertainties from the $\chi^2 $ fit. The exact contributions of statistical and systematic uncertainty to each fit in each \redmagic and \maglim bin are shown in Tables \ref{table:1paramerrors}-\ref{table:2paramerrors} in Appendix \ref{sec:errorbreakdown}.

\begin{table}
\begin{center}
    \begin{tabular}{|c|c|c|}
      \hline
      \multicolumn{3}{|c|}{\redmagic Results (1-parameter)} \\
      \hline
      Redshift Bin & $\Delta z$ & $\chi^2$ (points) \\
      \hline
      1: $z_{\text{ph}} \in [0.15,0.35]$ & 0.006 $\pm$ 0.004 & 6.81 (13)  \\   
      \hline
      2: $z_{\text{ph}} \in [0.35,0.5]$ & 0.001 $\pm$ 0.003 & 10.03 (11)  \\   
      \hline
      3: $z_{\text{ph}} \in [0.5,0.65]$ & 0.004 $\pm$ 0.003 & 7.32 (13)  \\   
      \hline
      4: $z_{\text{ph}} \in [0.65,0.8]$ & -0.002 $\pm$ 0.005 & 19.92 (16)  \\   
      \hline
      5: $z_{\text{ph}} \in [0.8,0.9]$ & 0.020 $\pm$ 0.010 & 69.35 (18)  \\   
      \hline
    \end{tabular}
  \caption{\redmagic clustering redshift results. $\Delta z$ is the shift that makes the photo-$z$ prediction from \redmagic match the mean of the clustering measurements. Statistical errors are from DES-reference cross-correlation and reference auto-correlations. Systematic errors are a combination of errors on the power law fit to the DES auto-correlation (Equation \ref{powerlaw}) and a 0.0025 method error from Section \ref{sec:simtests}. The bias of +0.0007 from that Section is also applied.}
  \label{table:redmagic}
\end{center}
\end{table}

\begin{table}
\begin{center}
    \begin{tabular}{|c|c|c|c|}
      \hline
      \multicolumn{4}{|c|}{\redmagic Results (2-parameter)} \\
      \hline
      Redshift Bin & $\Delta z$ & $s$ & $\chi^2$ (points) \\
      \hline
      1: $z_{\text{ph}} \in [0.15,0.35]$ & 0.007 $\pm$ 0.005 & 0.975 $\pm$  0.043 & 5.90 (13)\\   
      \hline
      2: $z_{\text{ph}} \in [0.35,0.5]$ & -0.002 $\pm$ 0.005 & 1.015 $\pm$ 0.045 & 7.14 (12)\\   
      \hline
      3: $z_{\text{ph}} \in [0.5,0.65]$ & 0.003 $\pm$ 0.005 & 1.017 $\pm$ 0.048 & 6.99 (11) \\   
      \hline
      4: $z_{\text{ph}} \in [0.65,0.8]$ & -0.002 $\pm$ 0.006 & 1.051 $\pm$ 0.065 & 17.54 (16) \\   
      \hline
      5: $z_{\text{ph}} \in [0.8,0.9]$ & -0.007 $\pm$ 0.006 & 1.230 $\pm$ 0.066 & 16.74 (18)\\   
      \hline
    \end{tabular}
  \caption{\redmagic clustering redshift results for a 2-parameter fit. $\Delta z$ is the shift parameter, and $s$ is the stretch parameter. Each is fit by changing the photo-$z$ distribution to match the clustering data points with a $\chi^2$ fit. Uncertainty comes from statistical uncertainty of the fit, which includes contributions from the power law fit uncertainty (Equation \ref{powerlaw}), and the `method error' in Section \ref{sec:47}.}
  \label{table:redmagic_2param}
\end{center}
\end{table}

\subsection{\redmagic Results}
\label{sec:redmagicresults}

\begin{figure*}
\begin{center}
\includegraphics[width=1.0 \textwidth]{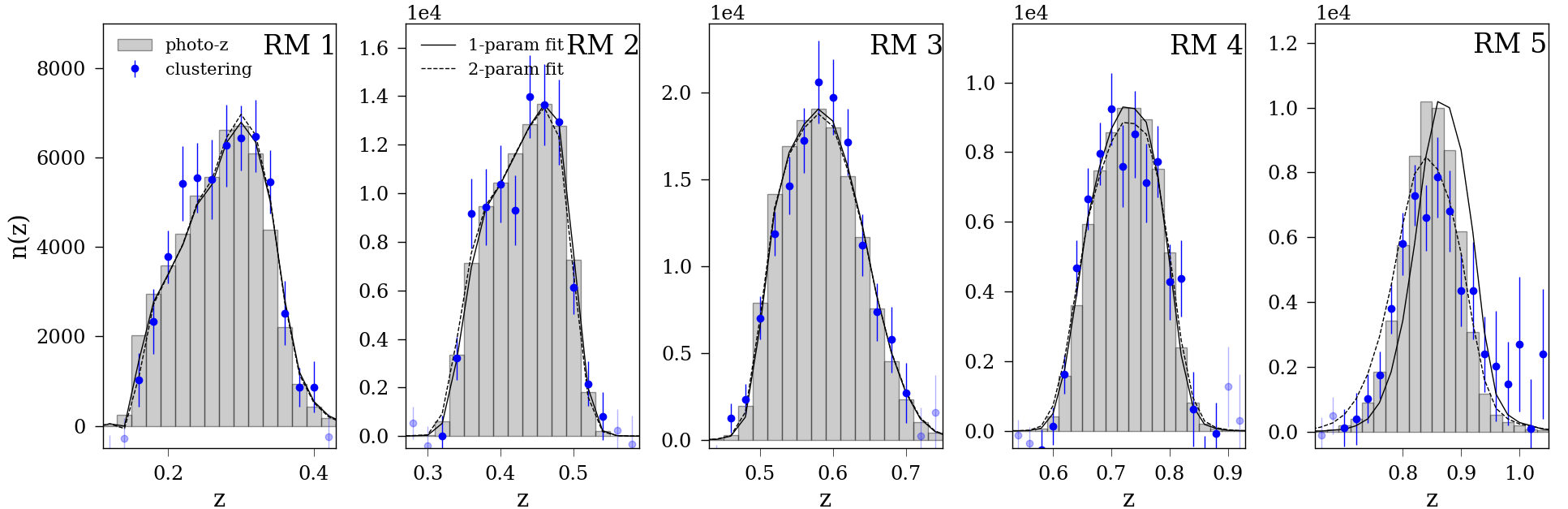}
\end{center}
\caption{The clustering redshift measurements for the five \redmagic bins. The photo-$z$ prediction comes from the \redmagic algorithm itself. The dark blue data points indicate the range where clustering and photo-$z$ are compared to find the best fit shift, $\Delta z$. The gray points are outside this range and not used. Error bars only reflect statistical errors from the cross-correlation of DES and reference (BOSS/eBOSS) galaxies, and the auto-correlation of the reference galaxies. Best fit parameters are given in Tables \ref{table:redmagic}-\ref{table:redmagic_2param}.}
\label{fig:redmagicresults}
\end{figure*}

\begin{figure*}
\begin{center}
\includegraphics[width=1.0 \textwidth]{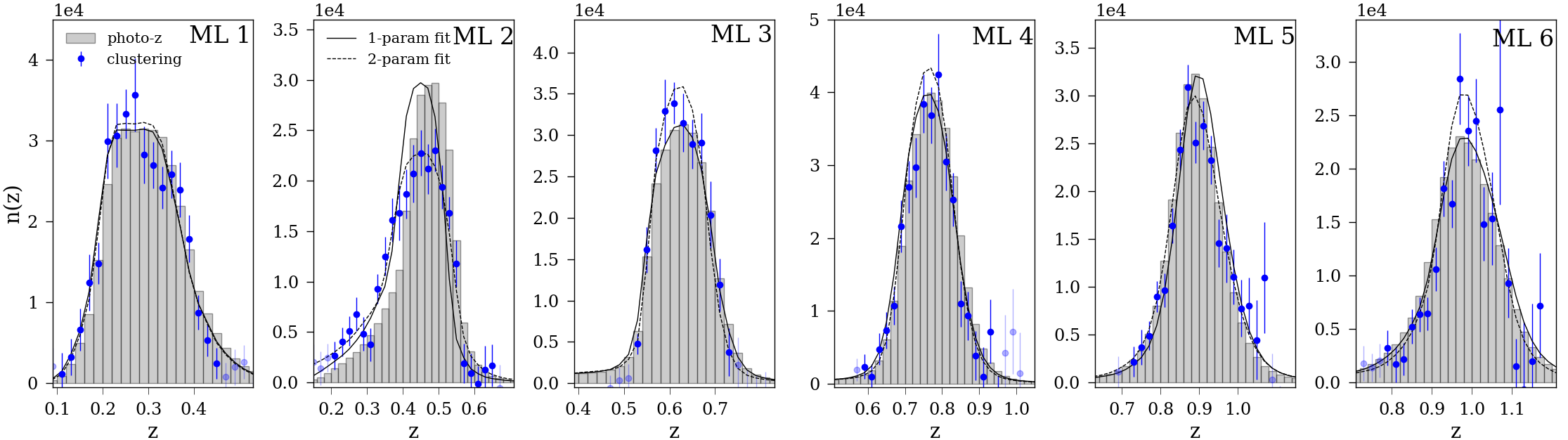}
\end{center}
\caption{The clustering redshift measurements for the \maglim sample. The photo-$z$ prediction comes from the summation of the entire \dnff probability distribution function for each galaxy. See Figure \ref{fig:redmagicresults} for more details about the data points. Best fit parameters are given in Tables \ref{table:maglim}-\ref{table:maglim_2param}.}
\label{fig:maglimresults}
\end{figure*}

\begin{figure}
\begin{center}
\includegraphics[width=0.5 \textwidth]{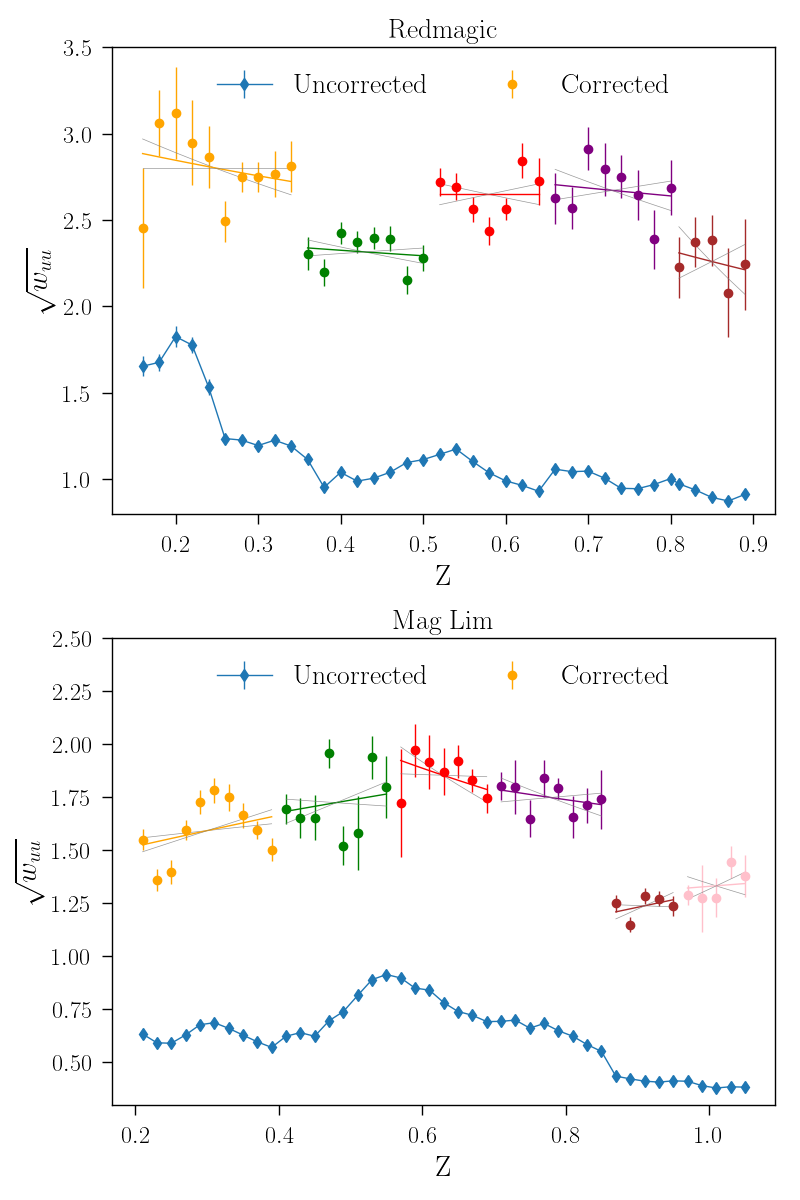}
\end{center}
\caption{Auto-correlations of the DES lens samples, \redmagic and \maglim. Shown is the square root of the auto-correlation, $\sqrt{\bar{w}_{\text{uu,pz}}(z_j)}$, listed as uncorrected. The corrected data points are $\sqrt{\bar{w}_{\text{uu,pz}}(z_j) \sigma_j}$}, to undo the influence of photo-$z$ scatter on the auto-correlations. The eleven different tomographic bins are differentiated by color. The solid lines in color are the best power law fit to the points in a given bin (Equation \ref{powerlaw}). The gray lines indicate the $1\sigma$ range of the power law fit. We note that due to normalization of the $n(z)$'s, the amplitudes of these measurements are not important, only the change across a single tomographic bin.
\label{fig:desauto}
\end{figure}

Our clustering redshift estimates for the \redmagic sample are shown in Figure \ref{fig:redmagicresults} and Tables \ref{table:redmagic}-\ref{table:redmagic_2param}. In the first four bins of Figure \ref{fig:redmagicresults}, we see a generally good agreement in the shapes and means of the clustering and photometric redshift distributions. The first four bins have relatively small biases ($\leq 0.006$), and are within $1.5$ standard deviations of zero bias for the 1-parameter fit. The fifth bin has a more obvious difference in shape between the clustering and photo-$z$ distributions, with a high-$z$ tail in the clustering redshift distribution. Driven by this tail, a large shift parameter of $0.02$ is needed to match the means of the distributions in the 1-parameter fit. 

There are different possible metrics in selecting whether the 1- or 2- parameter model should be used for the DES cosmology analysis. In a cosmology analysis, every extra parameter allowed to vary will typically reduce the constraining power of the experiment. Thus, there is a benefit to having a simpler model if it is accurate enough. One possible metric for deciding whether a 1- or 2-parameter model is sufficient is whether the 2-parameter model prefers $s \neq 1$. In Table \ref{table:redmagic_2param}, we see that the first four bins all are well fit by $s=1$, suggesting that a 2-parameter fit may be unnecessary. Bin 5, in contrast,  prefers $s \neq 1$ at $>3\sigma$ confidence.

Another metric could be to use the goodness of fit, which can be assessed with the $\chi^2$ values listed in Tables \ref{table:redmagic}-\ref{table:redmagic_2param}. We see that the \redmagic 1-parameter fits in the first four bins are all close to a reduced $\chi^2$ of 1. As expected, the fifth bin exhibits a very poor fit. We do see somewhat lower $\chi^2$ values in the 2-parameter fits for the first four bins, but at the cost of a second parameter (s), and increased errors on the $\Delta z$. In the fifth bin though, the 2-parameter fit is much better, with a reduced $\chi^2$ near 1. We note that though the fit may not look quite that good in Figure \ref{fig:redmagicresults}, the off-diagonal terms of the covariance between the $n(z)$ data points has a strong effect on this particular fit.

For the cosmology analysis in \cite{y3-3x2ptkp}, the general strategy is to add complexity to models only if it is necessary to not bias the cosmological results. With this strategy in mind, we show another test in assessing whether the 1- or 2-parameter models are needed in Appendix \ref{appendix:photozvalidation}. There, we show MCMC chains from a simulated cosmology analysis approximating the analysis in \cite{y3-3x2ptkp}. Chains are run with fixed cosmology but different redshift inputs, to assess whether the redshift modeling is sufficient to not bias cosmological results.

In Figure \ref{fig:chainsfull} in Appendix \ref{appendix:photozvalidation}, we show simulated results for $\Omega_{\text{m}}, \sigma_8$ and the galaxy bias in each of the five \redmagic bins for four different redshift inputs. The different inputs are: the clustering-redshift results directly, a multi-gaussian fit very closely matching the clustering-redshifts but somewhat smoother, and the 1- and 2- parameter fits listed in Tables \ref{table:redmagic}-\ref{table:maglim_2param}. We find in this test that the cosmological parameters are similar in all cases. However, the galaxy bias recovered in the fifth \redmagic bin is offset from the more direct clustering fits. This suggests that the poor fit of the 1-parameter model in the fifth bin will give biased results for the galaxy bias if the clustering-redshifts results are accurate.

Based on these two tests, the 1-parameter fits for the first four \redmagic bins, and the 2-parameter fit for the fifth bin were chosen as the fiducial models for \cite{y3-3x2ptkp}. In order to be conservative with the fifth bin, the uncertainty on $\Delta z$ was increased from 0.007 to 0.010, to match the uncertainty from the 1-parameter fit. We investigate the fifth \redmagic bin clustering results in more detail in Section \ref{sec:magnification}, specifically checking whether the high-$z$ tail could be explained by magnification. As shown there, we conclude it cannot be. We also note that we measured \redmagic bin 5 at the larger clustering scale range, 0.5-4 Mpc, which we use for two of the \maglim bins. With these scales, this high-$z$ tail for \redmagic bin 5 remains, with marginally smaller error bars.

\begin{table}
\begin{center}
    \begin{tabular}{|c|c|c|c|c|}
      \hline
      \multicolumn{3}{|c|}{\maglim Results (1-parameter)} \\
      \hline
      Redshift Bin & $\Delta z$ & $\chi^2$ (points) \\
      \hline
      1: $z_{\text{ph}} \in [0.2,0.4]$ & -0.010 $\pm$ 0.004 & 23.96 (18) \\   
      \hline
      2: $z_{\text{ph}} \in [0.4,0.55]$ & -0.028 $\pm$ 0.006 & 91.91 (23) \\   
      \hline
      3: $z_{\text{ph}} \in [0.55,0.7]$ & -0.004 $\pm$ 0.004 & 12.64 (11) \\   
      \hline
      4: $z_{\text{ph}} \in [0.7,0.85]$ & -0.010 $\pm$ 0.005 & 22.61 (19) \\   
      \hline
      5: $z_{\text{ph}} \in [0.85,0.95]$ & 0.013 $\pm$ 0.007 & 44.34 (18) \\   
      \hline
      6: $z_{\text{ph}} \in [0.95,1.05]$ & 0.009 $\pm$ 0.016 & 36.61 (20) \\   
      \hline
    \end{tabular}
  \caption{\maglim sample clustering redshift results. $\Delta z$ is the shift that makes the photo-$z$ prediction from \dnff match the mean of the clustering measurements. See Table \ref{table:redmagic} for further comments on uncertainty column sources.}
  \label{table:maglim}
\end{center}
\end{table}

\begin{table}
\begin{center}
    \begin{tabular}{|c|c|c|c|}
      \hline
      \multicolumn{4}{|c|}{\maglim Results (2-parameter)} \\
      \hline
      Redshift Bin & $\Delta z$ & $s$ & $\chi^2$ (points) \\
      \hline
      1: $z_{\text{ph}} \in [0.2,0.4]$ & -0.009 $\pm$ 0.007 & 0.975 $\pm$  0.062 & 24.27 (18) \\   
      \hline
      2: $z_{\text{ph}} \in [0.4,0.55]$ & -0.035 $\pm$ 0.011 & 1.306 $\pm$ 0.093 & 22.94 (23) \\   
      \hline
      3: $z_{\text{ph}} \in [0.55,0.7]$ & -0.005 $\pm$ 0.006 & 0.870 $\pm$ 0.054 & 6.55 (11) \\   
      \hline
      4: $z_{\text{ph}} \in [0.7,0.85]$ & -0.007 $\pm$ 0.006 & 0.918 $\pm$ 0.051 & 21.96 (19) \\   
      \hline
      5: $z_{\text{ph}} \in [0.85,0.95]$ & 0.002 $\pm$ 0.007 & 1.080 $\pm$ 0.067 & 25.79 (18) \\  
      \hline
      6: $z_{\text{ph}} \in [0.95,1.05]$ & 0.009 $\pm$ 0.008 & 0.845 $\pm$ 0.073 & 36.59 (20) \\  
      \hline
    \end{tabular}
  \caption{\maglim clustering redshift results for a 2-parameter fit. $\Delta z$ is the shift parameter, and $s$ is the stretch parameter. See See Table \ref{table:redmagic_2param} for further comments on the fit and uncertainties. A typo in bin 6 from the original version of this paper has been corrected.}
  \label{table:maglim_2param}
\end{center}
\end{table}

\subsection{\maglim Sample Results}
\label{sec:maglimresults}

Our clustering redshift estimates for the \maglim sample are shown in Figure \ref{fig:maglimresults} and Tables \ref{table:maglim}-\ref{table:maglim_2param}. We can see that there are significantly different shapes of the clustering and photo-$z$ distributions in multiple bins.  It was expected that this larger sample of fainter galaxies would have larger photo-$z$ biases than \redmagic.

It is again important to analyze whether a 1- or 2-parameter fit will be more appropriate for the cosmology analyses.
For the metric of checking whether the 2-parameter fit is consistent with $s=1$, we find in Table \ref{table:maglim_2param} only one bin where this is clearly the case (bin 1). In examining the $\chi^2$ fits for the 1- and 2-parameter fits in Tables \ref{table:maglim}-\ref{table:maglim_2param}, we see a very strong preference for the 2-parameter fit in bin 2 and bin 5, though it is less clear if the 2-parameter fit is necessary for the other bins by this metric.

The final decision on whether 1 or 2 parameters are needed for the \maglim fits is determined by the procedures in \cite{y3-3x2ptkp} and \cite{y3-generalmethods}, which focus on whether there will be biases in the cosmological analysis if the simpler model is used. These tests for the \maglim redshifts, analogous to the \redmagic tests in Appendix \ref{appendix:photozvalidation}, are shown in \cite{y3-2x2ptaltlensresults}, Figure 6. There it is shown that the galaxy bias in $\maglim$ bins 2-6 using the 1-parameter fits of (this paper's) Table \ref{table:maglim} are offset from what the estimates would be inputting the clustering redshift measurements directly. This would mean that the 1-parameter fits may bias the galaxy bias measurements. Thus, it was decided to use 2-parameter fits for all of the \maglim bins.

We note that our initial measurements of the two highest redshift bins for the \maglim sample were very noisy, mainly due to the low number of eBOSS objects to correlate with.  Our final analysis for bins 5 and 6, shown in Figure \ref{fig:maglimresults}, used the clustering redshift method on scales 0.5-4.0 Mpc, while the rest of this work used 0.5-1.5 Mpc. At these noisier high redshifts, we found including more scales improved the signal significantly, reducing total uncertainty on $\Delta z$ by about $50 \%$ and $80 \%$ for bins 5 and 6 respectively (for the 1-parameter fit). We tested other bins at these scale ranges and found negligible differences in other cases. Since we weight the smallest scales most, where more information is expected, this lack of change in most cases is unsurprising. 

We also checked the $z=1.06-1.08$ clustering data point, which stands out in bin 5, and somewhat in bin 6. An isolated peak like this appears strange and perhaps anomalous. We tested several things to try to find an issue with the clustering data, but could find none. It was not found in \redmagic when we extended the redshift range. It was not found in auto-correlations, or cross-correlations of the eBOSS ELG and QSO samples with each other. It was found in cross-correlations of \maglim with either the ELG or QSO alone. Changing the binning, the number of jackknife patches, or splitting up the data into different large regions did not remove the signal. We also tried not using either \maglim or eBOSS weights, and also estimators that would not use one of the sample's randoms. In all cases, the signal persisted. The signal is also much too large for magnification, which should not produce a sharp change in redshift anyway. We conclude that the signal is real, either some statistical fluctuation, or some interloping population of galaxies that got into the \maglim cuts.

\subsection{Comments on Redshift Impacts to Cosmological Analysis}
\label{sec:goodnessoffit}

The redshift parameters and choices described in this section for each \redmagic and \maglim were made before unblinding the cosmological results shown in \cite{y3-3x2ptkp}. In assessing those results post-unblinding, there were some redshift related tests worth noting.

One of the noteworthy issues discussed in \cite{y3-3x2ptkp} is the apparent disagreement of the \redmagic galaxy clustering plus galaxy-galaxy lensing results and the cosmic shear results. This is discussed in detail in \cite{y3-3x2ptkp}. One of the first tests done was to see if using the 2-parameter fits from Table \ref{table:redmagic_2param} in all \redmagic redshift bins could alleviate this issue. It did not. The resulting chains with the 2-parameter redshift model had only slightly larger contours and there was still significant inconsistency in the data.

Another significant test for understanding the \redmagic inconsistency was dropping different redshift bins. Most notably for this  work, the 5th \redmagic bin was shown to have little impact on the cosmological results and the inconsistency. Therefore, though the fifth \redmagic bin is clearly the one with the greatest redshift uncertainties, it cannot be driving the inconsistencies in the \redmagic results.

For the \maglim sample, a similar issue of significant mismatch between the galaxy-galaxy lensing+galaxy clustering amplitude and cosmic shear amplitude was found in Bins 5 and 6, as discussed in \cite{y3-2x2ptaltlensresults}. The measurements in the first four bins were internally consistent though and were used for the fiducial results in \cite{y3-3x2ptkp}. It is unclear at this time what the issue in these high redshift bins is. In \cite{y3-2x2ptaltlensresults} Appendix A, tests seemed to indicate that the problem is more with galaxy-galaxy lensing than with galaxy clustering. Based on this, it is unlikely errors in the lens redshifts are driving this tension. The galaxy clustering measurements will depend much more on the lens redshifts, particularly the width parameter.

\section{Magnification}
\label{sec:magnification}

In this section, we calculate whether magnification may significantly be affecting our results. Clustering redshifts are known to be affected by magnification effects (\citealt{choi16}, \citealt{gatti18}) which become significant in the tails of the redshift distribution where the normal clustering signal is small. Our cut of the tails at $2.5 \sigma$ from the peak of the clustering $n(z)$ should remove most of the redshift range where magnification effects are significant in each bin. However, some of our results do include a fairly large signal in the tails. Notably, the high-$z$ tail in \redmagic bin 5 is large enough to be within this cut and not removed.

We calculate a theory prediction for magnification in our clustering redshift measurements. Specifically, we will calculate the theoretical signal in the \redmagic bin 5 measurements to assess whether the high-$z$ tail is magnification-induced, or is real evidence for a photo-$z$ bias. We first calculate the strength of the galaxy clustering signal between the two samples (\citealt{choi16}):

\begin{multline}
\label{ggtheory}
 w(\theta)^{\text{gg}}_{\text{ur}}=b_{\text{u}} b_{\text{r}} \int n_{\text{u}}(z(\chi)) n_{\text{r}}(z(\chi)) \frac{dz}{d\chi} \frac{dz}{d\chi} d\chi \\ \times \int \frac{k}{2 \pi} P(k,z(\chi)) J_0(\chi k \theta) dk   
\end{multline}

\noindent where $\theta=\frac{r}{\chi(z)}$ as described in Section \ref{sec:methods}, $b$ is the galaxy bias for the unknown (DES) and reference (eBOSS) samples, $n$ is the redshift distribution of each sample, $\chi$ is the comoving distance, $k$ is the wavenumber, $P(k)$ is the matter power spectrum, and $J_0$ is the zeroth order Bessel function. 

We also calculate the strength of the magnification signal, specifically the signal from foreground \redmagic galaxies lensing eBOSS galaxies (the magnification effect that could be found on the high-$z$ end):

\begin{multline}
\label{gmutheory}
w(\theta)^{\text{g} \mu}_{\text{ur}}=b_{\text{u}} (\alpha-1) \int n_{\text{u}}(z) K(\chi) \frac{dz}{d\chi}  d\chi \\ \times \int \frac{k}{2 \pi} P(k,z(\chi)) J_0(\chi k \theta) dk .
\end{multline}

\noindent $K(\chi)$ is the lensing kernel:

\begin{equation}
\label{lensingkernel}
K(\chi)=\frac{3 H_0^2 \Omega_{\text{m}}}{c^2}\frac{\chi}{a} \int_{\chi}^{\infty} n_{\text{r}}(z) \frac{dz}{d \chi} \frac{\chi'-\chi}{\chi'} d \chi'
\end{equation}

\noindent where $H_0$ is the Hubble constant, $\Omega_{\text{m}}$ is the matter density parameter, $c$ is the speed of light, and $a=1/(1+z)$ is the scale factor of the Universe. In Equation \ref{gmutheory}, $\alpha$ is the slope of the magnitude counts for, in this case, the eBOSS sample: 

\begin{equation}
\label{alpha}
\alpha(m)=2.5 \frac{d \text{log}_{10} n(>m)}{dm}
\end{equation}

\noindent where $m$ is the limiting magnitude of the galaxy sample (\citealt{choi16}). For eBOSS ELG (the main tracer used for the redshift range of \redmagic bin 5), we calculate a value of $\alpha=2.71$ at its limiting g-band magnitude, $m=22.825$ mag (\citealt{raichoor17}).


\begin{figure}
\begin{center}
\includegraphics[width=0.5 \textwidth]{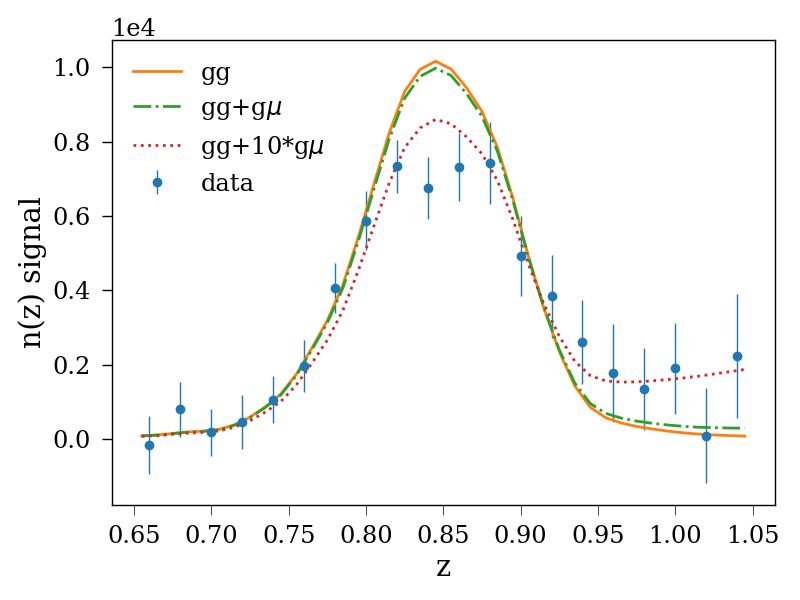}
\end{center}
\caption{Estimates of the  $n(z)$ signal for \redmagic bin 5 ($z_{\text{ph}} \in [0.8,0.9]$). For the purpose of comparison, the clustering data points are only from the cross-correlation of \redmagic and eBOSS, making them slightly different from Figure \ref{fig:redmagicresults}. In orange (gg), is the theoretical prediction for this cross-correlation due to clustering alone (Equation \ref{ggtheory}). In green (gg+$g\mu$) is the theoretical prediction from galaxy clustering and correlations between \redmagic galaxies and magnification effects on eBOSS galaxies (Equations \ref{ggtheory}-\ref{gmutheory}). In red (gg+10*g$\mu$) is a theory prediction with galaxy clustering and a ten times larger amplitude prediction from the galaxy-magnification signal. Each of the theory predictions uses the photometric $n(z)$ prediction for the bin as input. Effects of magnification do not seem to be large enough in theory to account for the excess signal at high redshifts in this bin. There is no theoretical motivation for a factor of ten error in magnification predictions.}
\label{fig:magnification}
\end{figure}

We ignore the terms where eBOSS galaxies could magnify background \redmagic galaxies. Since the eBOSS galaxies have spectroscopic redshifts, we know the galaxies being correlated in the high-$z$ tail are at $z \sim 1$. Thus, for eBOSS galaxies to be magnifying \redmagic galaxies, there would need to be a significant number of very high redshift outliers in the \redmagic population. On the other hand, \redmagic galaxies around $z \sim 0.8$ in this fifth bin could plausibly magnify eBOSS galaxies at $z=1$ where the excess is seen. We also ignore the magnification-magnification term which should be negligible, particularly for two galaxy samples not widely separated in redshift space (see e.g., \citealt{duncan14}). To do the calculations, we assume for $n_{\text{u}}(z)$, the photometric redshift distribution estimate as an input, and use Planck 2015 flat-$\Lambda$CDM cosmological parameters including external data (\citealt{planck15}). These parameters include $H_0=67.74 \ \text{km/sec/Mpc}$ and $\Omega_{\text{m}}=0.3089$. The power spectrum $P(k)$ is calculated using the Boltzmann code in CAMB (\citealt{camb1}, \citealt{camb2}) with Halofit (\citealt{halofit}) used to calculate nonlinear clustering effects. Minor deviations on the input redshift distribution or cosmology do not change qualitatively the results. Since the galaxy clustering and galaxy-magnification equations both contain a factor of $b_{\text{u}}$, it effectively cancels in a normalized clustering redshift calculation, so we set it to 1. We ignore the galaxy bias evolution, which will only produce small changes. We also set $b_{\text{r}}=1$, which is close to other studies of eBOSS ELGs which indicate $b_{\text{r}}=1.3$ (\citealt{guo2019}).

We show the results of our theory calculations for the clustering-clustering and clustering-magnification terms for \redmagic bin 5 as well as the measurements in Figure \ref{fig:magnification}. The results show that the magnification signal is far too small to explain the measured high-$z$ excess in the bin. We show that the magnification term would need roughly a factor of 10 increase to match the data, so small errors in e.g., the $\alpha$ calculated in Equation \ref{alpha} could not explain the excess. 

We perform similar tests on other bins and consistently see a theoretical magnification signal that is negligible to our results. This is mainly due to two factors, one is the relatively narrow redshift bins used, reducing the amplitude of the lensing kernel (Equation \ref{lensingkernel}) between the DES galaxies and eBOSS galaxies. For example, with wider bins, one could have $z \sim 0.6$ DES galaxies lensing $z=1$ eBOSS galaxies, which would have a larger lensing kernel than $z=0.8$ DES galaxies. The second factor is our procedure of cutting the tails where the signal is small as described in Section \ref{sec:33}. This cuts out the regions with both highest magnification signal and lowest clustering signal.

\begin{figure}
\begin{center}
\includegraphics[width=0.5 \textwidth]{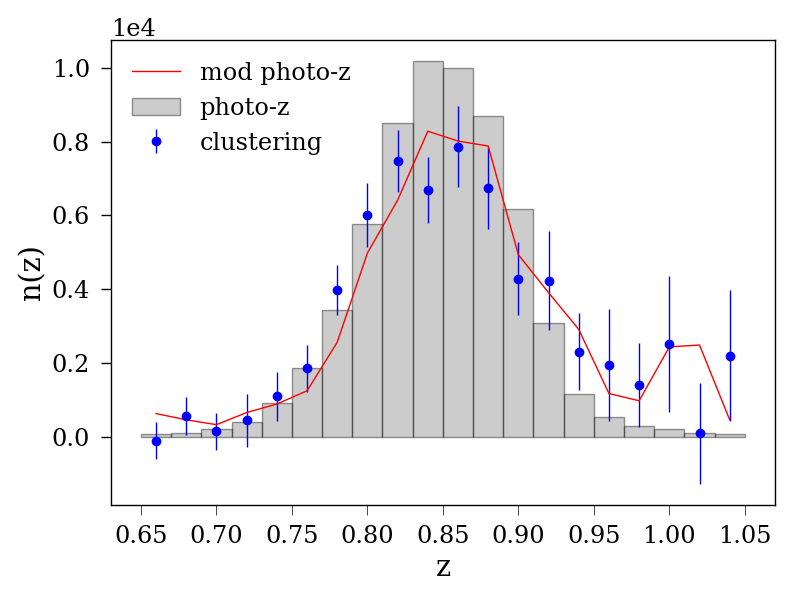}
\end{center}
\caption{In red, the extrapolated $n(z)$ from spectroscopic redshifts in \redmagic bin 5. The extrapolation is based on the ratio of galaxies in each $dz=0.02$ bin by spec-$z$ and by `ZMC', a draw from the \redmagic redshift distribution function in the subset of galaxies with spec-$z$ measurements. This ratio is then applied to the full \redmagic sample (the portion overlapping eBOSS) to give this extrapolated prediction. We see it matches the clustering results well, giving some evidence that the high-$z$ tail is physical, rather than a clustering systematic.}
\label{fig:speccheck}
\end{figure}

We note that there is also other evidence that the high-$z$ excess in \redmagic bin 5 is a true photo-$z$ bias. In the DES/eBOSS overlap area (mostly overlapping the region known as Stripe 82), about $3\%$ of the \redmagic bin 5 galaxies have spectroscopic redshifts. In this subsample, a similar high-$z$ excess in the spectroscopic $n(z)$ is seen compared to the photometric estimate. We extrapolate a prediction for the $n(z)$ based on this spectroscopic subsample in Figure \ref{fig:speccheck}. For this, we take the photometric $n(z)$ for the whole Stripe 82 region, and multiply that by a correction factor for each micro-bin ($dz=0.02$ in size) where the correction factor is $n_{\text{spec}}/n_{\text{pz}}$ as measured in the $3 \%$ subsample with spectroscopic redshifts. We see that there is good agreement with the clustering redshifts signal.

We caution that comparing photo-$z$ errors on a sample with spectroscopic redshifts to a full sample is not generally reliable. Mismatches are seen when applying this procedure on other bins. This is generally due to the fact that the subsamples with spectroscopic measurements tend to be brighter than the full samples, and will typically have smaller photo-$z$ errors. Since this particular error is seen in the brighter subsample of bin 5 though, it is more likely to be present in the full sample than the opposite case (i.e., assuming a lack of error in the bright sample extrapolates well to the full sample). Between this minor evidence and the magnification calculations, we conclude the high-$z$ excess in the last \redmagic bin is likely real. We also note that in early versions of the Year-3 \redmagic catalog, we saw more significant biases when the 5th bin extended to $z=0.95$, also suggesting the \redmagic algorithm is encountering issues at high redshift.

\section{Summary}
\label{sec:summary}

In this work, we present clustering redshift measurements of two DES lens samples, \redmagic and \maglim (Figures \ref{fig:redmagicresults}-\ref{fig:maglimresults}). These measurements inform the redshift models for the analyses in \cite{y3-3x2ptkp} and related papers. Our results are bolstered by the large number of spectroscopic galaxies available for this measurement from the BOSS and eBOSS galaxy clustering catalogs, and their several hundred $\text{deg}^2$ overlap with DES. We generally find small biases ($|\Delta|<0.01$) for the photometric redshift predictions of these samples (Tables \ref{table:redmagic}-\ref{table:maglim_2param}). Our results suggest the shape of the \redmagic photo-$z$ distributions in particular are very accurate. The fainter, larger \maglim galaxy sample had more significant differences in shape when comparing the photo-$z$ and clustering distributions, suggesting the need for a 2-parameter fit for calibration.  We were able to constrain the mean redshifts, in the form of the bias parameter $\Delta z$, of the different bins to a precision of typically around 0.005 when doing 1-parameter fits. Our uncertainties on the mean redshifts were only marginally larger when doing 2-parameter fits.

We tested our methodology in simulations (Section \ref{sec:simtests}), including a new method of calibrating the galaxy bias systematic of the `unknown'  sample, the DES galaxies. This new method, involving cross-correlations on smaller redshift ranges, is made possible by the large number of spectroscopic tracers we had compared to previous work. We found the systematic errors on our method, beyond the statistical errors we account for in all of the auto- and cross-correlations, to be quite small for calculating a single photo-$z$ bias (Figure \ref{fig:wztests} and Table \ref{table:test_results}). We also tested our ability to constrain the width of a redshift distribution. In this case, we found more significant `method' errors not accounted for by the initial statistics. Despite this, our results suggest we can constrain the width to around $4-7\%$ for most samples. In Appendix \ref{sec:densitydependenceappendix}, we show some tests that suggest that density dependence of the method may increase these uncertainties up to around $11 \%$, but this would have negligible impact on the DES cosmological results.

In one \redmagic bin, we investigate specifically if a high-redshift tail in our results can be explained by magnification, finding it cannot. Our procedure of cutting the tails of the clustering redshift distribution, and the relatively narrow redshift bins for the DES lens samples in general, make magnification a negligible effect for our work.

We also show for that particular \redmagic bin in Appendix \ref{appendix:photozvalidation}, that a 2-parameter fit to the clustering data should not bias our cosmological results (while a 1-parameter fit could at least bias the galaxy bias estimates). A similar analysis for \maglim is shown in \cite{y3-2x2ptaltlensresults}, where it is found that most of the \maglim bins need a 2-parameter fit as well to not bias cosmological results. These tests show the importance of using a multi-parameter fit in calibrating photometric redshifts.

Our results provide important redshift constraints for our companion Dark Energy Survey Year 3 Results papers using \redmagic and \maglim galaxies for galaxy clustering and galaxy-galaxy lensing (\citealt{y3-3x2ptkp}, \citealt{y3-2x2ptbiasmodelling}, \citealt{y3-2x2ptaltlensresults}, \citealt{y3-galaxyclustering}, \citealt{y3-gglensing}, \citealt{y3-2x2ptmagnification}). This work also signifies some important steps in the progression of clustering redshift measurements, particularly in testing new methods to correct the galaxy bias systematic, and to constrain a width parameter of a redshift distribution. Future photometric and spectroscopic surveys will need to continue to develop the clustering redshifts technique as they push to higher redshifts.

\section*{Acknowledgements}
RC and KB acknowledge support from the U.S. Department of Energy, Office of Science, Office of High Energy Physics, under Award Numbers DE-SC0020278 and DE-SC0017647.

Funding for the DES Projects has been provided by the U.S. Department of Energy, the U.S. National Science Foundation, the Ministry of Science and Education of Spain, 
the Science and Technology Facilities Council of the United Kingdom, the Higher Education Funding Council for England, the National Center for Supercomputing 
Applications at the University of Illinois at Urbana-Champaign, the Kavli Institute of Cosmological Physics at the University of Chicago, 
the Center for Cosmology and Astro-Particle Physics at the Ohio State University,
the Mitchell Institute for Fundamental Physics and Astronomy at Texas A\&M University, Financiadora de Estudos e Projetos, 
Funda{\c c}{\~a}o Carlos Chagas Filho de Amparo {\`a} Pesquisa do Estado do Rio de Janeiro, Conselho Nacional de Desenvolvimento Cient{\'i}fico e Tecnol{\'o}gico and 
the Minist{\'e}rio da Ci{\^e}ncia, Tecnologia e Inova{\c c}{\~a}o, the Deutsche Forschungsgemeinschaft and the Collaborating Institutions in the Dark Energy Survey. 

The Collaborating Institutions are Argonne National Laboratory, the University of California at Santa Cruz, the University of Cambridge, Centro de Investigaciones Energ{\'e}ticas, 
Medioambientales y Tecnol{\'o}gicas-Madrid, the University of Chicago, University College London, the DES-Brazil Consortium, the University of Edinburgh, 
the Eidgen{\"o}ssische Technische Hochschule (ETH) Z{\"u}rich, 
Fermi National Accelerator Laboratory, the University of Illinois at Urbana-Champaign, the Institut de Ci{\`e}ncies de l'Espai (IEEC/CSIC), 
the Institut de F{\'i}sica d'Altes Energies, Lawrence Berkeley National Laboratory, the Ludwig-Maximilians Universit{\"a}t M{\"u}nchen and the associated Excellence Cluster Universe, 
the University of Michigan, the National Optical Astronomy Observatory, the University of Nottingham, The Ohio State University, the University of Pennsylvania, the University of Portsmouth, 
SLAC National Accelerator Laboratory, Stanford University, the University of Sussex, Texas A\&M University, and the OzDES Membership Consortium.

Based in part on observations at Cerro Tololo Inter-American Observatory at NSF's NOIRLab (NOIRLab Prop. ID 2012B-0001; PI: J. Frieman), which is managed by the Association of Universities for Research in Astronomy (AURA) under a cooperative agreement with the National Science Foundation.

The DES data management system is supported by the National Science Foundation under Grant Numbers AST-1138766 and AST-1536171.
The DES participants from Spanish institutions are partially supported by MINECO under grants AYA2015-71825, ESP2015-66861, FPA2015-68048, SEV-2016-0588, SEV-2016-0597, and MDM-2015-0509, 
some of which include ERDF funds from the European Union. IFAE is partially funded by the CERCA program of the Generalitat de Catalunya.
Research leading to these results has received funding from the European Research
Council under the European Union's Seventh Framework Program (FP7/2007-2013) including ERC grant agreements 240672, 291329, and 306478.
We  acknowledge support from the Brazilian Instituto Nacional de Ci\^encia
e Tecnologia (INCT) e-Universe (CNPq grant 465376/2014-2).

This manuscript has been authored by Fermi Research Alliance, LLC under Contract No. DE-AC02-07CH11359 with the U.S. Department of Energy, Office of Science, Office of High Energy Physics.

Funding for the Sloan Digital Sky Survey IV has been provided by the Alfred P. Sloan Foundation, the U.S. Department of Energy Office of Science, and the Participating Institutions. SDSS acknowledges support and resources from the Center for High-Performance Computing at the University of Utah. The SDSS web site is www.sdss.org.

SDSS is managed by the Astrophysical Research Consortium for the Participating Institutions of the SDSS Collaboration including the Brazilian Participation Group, the Carnegie Institution for Science, Carnegie Mellon University, Center for Astrophysics | Harvard \& Smithsonian (CfA), the Chilean Participation Group, the French Participation Group, Instituto de Astrofísica de Canarias, The Johns Hopkins University, Kavli Institute for the Physics and Mathematics of the Universe (IPMU) / University of Tokyo, the Korean Participation Group, Lawrence Berkeley National Laboratory, Leibniz Institut für Astrophysik Potsdam (AIP), Max-Planck-Institut für Astronomie (MPIA Heidelberg), Max-Planck-Institut für Astrophysik (MPA Garching), Max-Planck-Institut für Extraterrestrische Physik (MPE), National Astronomical Observatories of China, New Mexico State University, New York University, University of Notre Dame, Observatório Nacional / MCTI, The Ohio State University, Pennsylvania State University, Shanghai Astronomical Observatory, United Kingdom Participation Group, Universidad Nacional Autónoma de México, University of Arizona, University of Colorado Boulder, University of Oxford, University of Portsmouth, University of Utah, University of Virginia, University of Washington, University of Wisconsin, Vanderbilt University, and Yale University.

\section*{Data Availability}
The full Year 3 \redmagic and \maglim catalogs are available at this URL: \url{https://des.ncsa.illinois.edu/releases}. The subsets of these catalogs used in this manuscript will be made available upon reasonable request to the corresponding author.

\bibliographystyle{mnras_rc_arxiv}
\bibliography{wz_bibliography_20,des_y3kp}

\appendix
\section{Validating the Photometric Redshift Model}
\label{appendix:photozvalidation}

\begin{figure*}
\begin{center}
\includegraphics[width=1.0 \textwidth]{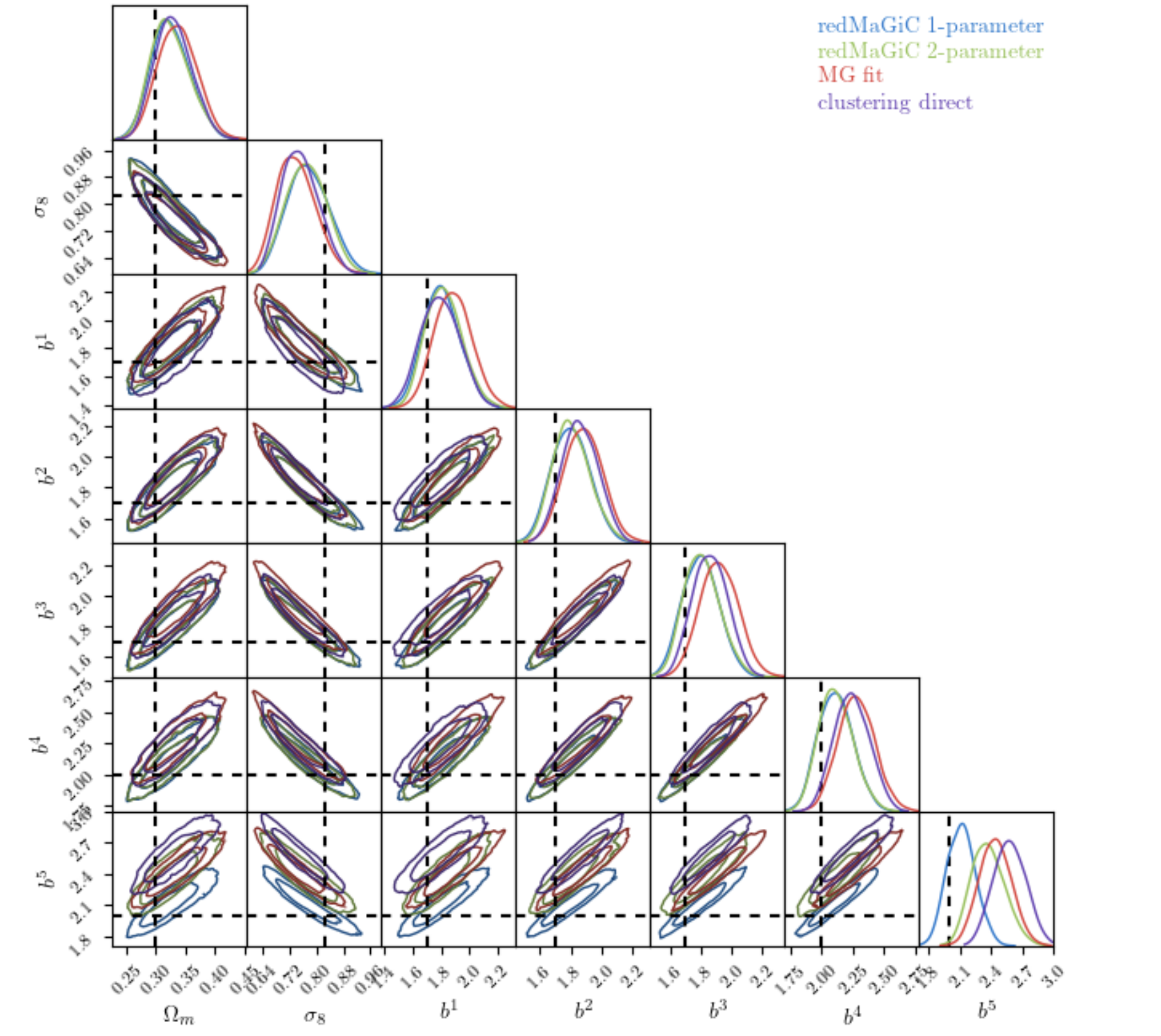}
\end{center}
\caption{Simulated cosmological constraints from galaxy clustering and galaxy-galaxy lensing with four different redshift distribution models for the \redmagic sample. The one-parameter contour uses the \redmagic photometric redshift distribution with shift parameters in all redshift bins from the results in this work (Table \ref{table:redmagic}). The two-parameter contour uses a two parameter shift and stretch model for the 5th \redmagic bin (Table \ref{table:redmagic_2param}). The MG fit contour uses a multi-Gaussian fit to the clustering redshift points. The clustering direct contour uses a spline fit to the clustering redshift points. The `true redshift' data vector for the simulation is the unshifted \redmagic photo-$z$ prediction. We can see that the 2-parameter model gives a better fit for the galaxy bias in bin 5 when compared to the predictions of the more exact fits to the clustering data, spline and multi-Gaussian.}
\label{fig:chainsfull}
\end{figure*}

Here, we show a test used to determine which photometric redshift model fits (1- or 2-parameters) sufficiently match the clustering redshift measurements such that it would not bias cosmological results. We want a photo-$z$ model that is flexible enough to agree with the clustering redshift points directly. We represent `clustering points directly' in two ways, one with a multi-Gaussian fit to the clustering data, and one with a spline. The spline will more exactly fit to the clustering data points, so we call it `clustering direct' in Figure \ref{fig:chainsfull}. The multi-Gaussian fit may be more realistic due to the spline overfitting to noise in the clustering data points. For the fiducial cosmology used in \cite{y3-generalmethods}, we calculate a noiseless galaxy clustering and galaxy-galaxy lensing data vector given the different redshift distributions. We ignore cosmic shear (the third part of the `3x2' measurement) since it does not use the lens galaxies. We then run MCMC chains on these measurements to infer cosmological and galaxy bias parameters. In these chains, all cosmology and intrinsic alignment parameters are allowed to vary and other nuisance parameters (such as the weak lensing source redshifts) are fixed. We follow the model validation analysis choices outlined in \cite{y3-generalmethods}. 

We ran chains with an input `true redshift' distribution matching the unshifted, unstretched \redmagic photo-$z$ prediction. We then ran chains with four different redshift models: a 1-parameter model using $\Delta z$ in all bins to shift the photo-$z$ predictions to match clustering, a two-parameter model that introduces a stretch parameter $s$ in the 5th \redmagic bin, a multi-Gaussian fit to the clustering data points, and a spline fit to the clustering data points. The results are shown in Figure \ref{fig:chainsfull}. In the first four \redmagic bins, we get significant overlap for the constraints on $\sigma_8$, $\Omega_{\text{m}}$ and the galaxy bias, $b$, with our 1-parameter model, and the redshift distributions that serve as a proxy for the clustering results. This signifies a good match between the clustering redshifts, and the 1-parameter shifted photometric redshifts.

We do see that for the 5th \redmagic bin, the galaxy bias is different comparing the one-parameter model contours to the multi-Gaussian and direct clustering fits. We can infer this means that a shifted \redmagic photo-$z$ distribution in this bin is still different enough from the clustering redshift distribution that they would produce statistically different amplitudes for galaxy clustering predictions in a given cosmology. The two-parameter contours show the constraints when the two-parameter fit is used in just the 5th bin. The addition of the stretch parameter, $s$, mitigates the discrepancy. With either the 1- or 2-parameter models, the cosmological constraints are unbiased compared to the clustering direct models. The specific $\sigma$ differences between the two-parameter model and the multi-Gaussian fit for $\Omega_{\text{m}}, \sigma_8$ and $b^5$ are 0.26, 0.39 and 0.31 respectively. The $\sigma$ differences between the two-parameter model and the `clustering direct' spline fit for $\Omega_{\text{m}}, \sigma_8$ and $b^5$ are 0.11, 0.24 and 0.95 respectively. Since the `clustering direct' and multi-Gaussian fits ignore the clustering uncertainties on shape, these values may effectively overestimate how much our model disagrees with the clustering data. 

\section{Breakdown of Uncertainty Contributions to Main Results}
\label{sec:errorbreakdown}

Here, we show the uncertainty contributions for our main results in Tables \ref{table:redmagic}-\ref{table:maglim_2param}. The calculations for overall uncertainty differ slightly between the 1- and 2-parameter cases. In the 1-parameter case, there are three contributions of uncertainty. The first is the statistical errors from the cross-correlations of the unknown and reference samples and the auto-correlations of the reference samples. These give an uncertainty on the mean redshift in Equation \ref{bothauto} with $w_{\text{uu}}$ set to 1. The second contribution is from doing the power law fit to the estimate of $w_{\text{uu}}$ in Equation \ref{powerlaw}. The uncertainty in the exponent $\gamma$ is translated to an uncertainty contribution on the mean redshift of a photometric bin. Finally, the third contribution is the systematic uncertainty derived in Table \ref{sec:simtests}.

For the 2-parameter fit, the $\chi^2$ fit for both $\Delta z$ and $s$ is done after the estimate and uncertainties of the power law fit to $w_{\text{uu}}$ (Equation \ref{powerlaw}) are already applied. Thus, the uncertainties in this fit encompass the first two contributions of the 1-parameter fit. We call this combination the statistical uncertainty. To this uncertainty, the systematic uncertainties from Table \ref{table:test_results_2param} are added. In each case of the 1- and 2-parameter uncertainty calculations, the contributions are added in quadrature. Each contribution for each redshift bin is shown in Tables \ref{table:1paramerrors}-\ref{table:2paramerrors}.

\begin{table}
\begin{center}
    \begin{tabular}{|c|c|c|c|c|}
      \hline
      \multicolumn{5}{|c|}{1-Parameter Uncertainty Contributions} \\
      \hline
      Redshift Bin & Total & Statistical & Power Law & Systematic \\
      \hline
      \redmagic Bin 1 & 0.004 & 0.0026 & 0.0011 & 0.0025  \\   
      \hline
      \redmagic Bin 2 & 0.003 & 0.0016 & 0.0006 & 0.0025  \\
      \hline
      \redmagic Bin 3 & 0.003  & 0.0018 & 0.0011 & 0.0025  \\ 
      \hline
      \redmagic Bin 4 & 0.005 & 0.0038 & 0.0015 & 0.0025  \\
      \hline
      \redmagic Bin 5 & 0.010  & 0.0089 & 0.0040 & 0.0025  \\   
      \hline
      \maglim Bin 1 & 0.004 & 0.0028 & 0.0015 & 0.0025  \\   
      \hline
      \maglim Bin 2 & 0.006 & 0.0036 & 0.0035 & 0.0025  \\   
      \hline
      \maglim Bin 3 & 0.004 & 0.0027 & 0.0018 & 0.0025  \\   
      \hline
      \maglim Bin 4 & 0.005 & 0.0036 & 0.0019 & 0.0025  \\   
      \hline
      \maglim Bin 5 & 0.011 & 0.0097 & 0.0044 & 0.0025  \\   
      \hline
      \maglim Bin 6 & 0.015 & 0.0127 & 0.0067 & 0.0025  \\   
      \hline
    \end{tabular}
  \caption{1-parameter fit uncertainty contributions for our main 1-parameter results in Tables \ref{table:redmagic} and \ref{table:maglim}.}
  \label{table:1paramerrors}
\end{center}
\end{table}

\begin{table}
\small\addtolength{\tabcolsep}{-2pt}
\begin{center}
    \begin{tabular}{|c|c|c|c|c|c|c|}
      \hline
      \multicolumn{7}{|c|}{2-Parameter Uncertainty Contributions} \\
      \hline
      Bin & Tot ($\Delta z$) & Stat ($\Delta z$) &  Syst ($\Delta z$) & Tot ($s$) & Stat ($s$) &  Syst ($s$) \\
      \hline
      RM 1 & 0.005 & 0.0025 & 0.0044 & 0.043 & 0.020 & 0.038  \\   
      \hline
      RM 2 & 0.005 & 0.0017 & 0.0044 & 0.045 & 0.023 & 0.038  \\
      \hline
      RM 3 & 0.005  & 0.0029 & 0.0044 & 0.0025 & 0.029 & 0.038  \\ 
      \hline
      RM 4 & 0.006 & 0.0035 & 0.0044 & 0.048 & 0.053 & 0.038  \\
      \hline
      RM 5 & 0.006 & 0.0041 & 0.0044 & 0.065 & 0.052 & 0.038  \\   
      \hline
      ML 1 & 0.007 & 0.0048 & 0.0044 & 0.062 & 0.049 & 0.038  \\   
      \hline
      ML 2 & 0.011 & 0.0097 & 0.0044 & 0.093 & 0.085 & 0.038  \\   
      \hline
      ML 3 & 0.006 & 0.0036 & 0.0044 & 0.054 & 0.039 & 0.038  \\   
      \hline
      ML 4 & 0.006 & 0.0033 & 0.0044 & 0.051 & 0.033 & 0.038  \\   
      \hline
      ML 5 & 0.007 & 0.0049 & 0.0044 & 0.067 & 0.056 & 0.038  \\   
      \hline
      ML 6 & 0.008 & 0.0064 & 0.0044 & 0.073 & 0.062 & 0.038  \\   
      \hline
    \end{tabular}
  \caption{2-parameter fit uncertainty contributions for our main 2-parameter results in Tables \ref{table:redmagic_2param} and \ref{table:maglim_2param}.}
  \label{table:2paramerrors}
\end{center}
\end{table}

\section{Alternative Assessments of Simulation Tests}
\label{sec:testsappendix}

In this Appendix, we describe a few variations on the evaluation of the tests on simulations Section \ref{sec:simtests}, focusing on the 1-parameter fit results in Table \ref{table:test_results}. One alternative would be to throw out a model having `biased' tests at all, and only fit for additional uncertainty (i.e., fit for $\omega$ in Equation \ref{chisquare}, and set $b=0$). We do evaluate the 2-parameter fit in this manner in Section \ref{sec:47}. In this case of no bias allowed, we would get uncertainty for Test 1, $\omega=0.0016$, and for Test 5, $\omega=0.0029$, for a combined uncertainty of 0.0033. For these calculations, we modify the degrees of freedom to be 5 instead of 4 in Equation \ref{chisquare}. This test would be a very small increase in method systematic uncertainty from the fiducial choice, 0.0025. 

We also address the question of whether it is appropriate that the six samples are treated as independent tests. While the six DES samples are different, there is significant redshift overlap between the \redmagic and magntiude-limited samples, thus the same simulated reference BOSS/eBOSS galaxies are used for multiple samples. With only six simulated samples, it is difficult to definitively prove correlation or lack thereof. Ideally, a large set of simulated galaxy samples, in different regions of the sky, could test this. Within our limited samples, we can do a few tests though.

If there are correlations in the measured biases between samples, we would expect them to be on the bins with the most overlap in redshift space. The pairs of samples with significant redshift overlap in Figures \ref{fig:simwz}-\ref{fig:wztests}  are [1,2], [3,4] and [5,6]. Analyzing only (mostly) non-overlapping redshift bins would likely have no correlations. We test this by performing again the calculations in Table \ref{table:test_results} using separately just the three \redmagic bins (samples 1,3,5) and \maglim bins (samples 2,4,6). The results for Tests 1 and 5 are shown in Table \ref{table:test_results_appendix}. For \redmagic alone, combining the Test 1 and Test 5 results in an overall bias of $b=0.0007$ and systematic uncertainty $\omega=0.0044$. For \maglim alone, the combined results give $b=0.0011$ and $\omega=0.0034$. These results are very similar to our fiducial values of $b=0.0007, \omega=0.0025$. In the most pessimistic case of total correlation between overlapping samples, using just the three \maglim samples changes our bias and uncertainty measurements by less than 0.001 each.

Although we have limited information in testing the hypothesis that the pairs of overlapping samples give correlated results, we can look at Figure \ref{fig:wztests}. If we naively looked for correlations in for example the Test 1 and Test 5 results between samples, we might conclude that Test 1 has correlations in samples 3 and 6, and Test 5 has correlations in samples 1 and 6. These samples are very unlikely to be correlated though since they cover different redshift ranges. This provides a bit more evidence that the tests are likely uncorrelated.

A simple $w(\theta)$ measurement of overlapping galaxy samples should certainly be correlated. However, it is not clear that the many $w(\theta)$ measurements in the full clustering redshift methodology, including the different angular bins in Equation \ref{lsequation}, and different redshift bins in e.g., Equation \ref{bothauto} should be correlated in their measurements of mean redshift. We leave a more explicit test of this for future work with a larger number of simulated samples.

We note that an evaluation of the 2-parameter fits in Figure \ref{fig:wztests2param} potentially indicates correlation in biases on the stretch parameter, unlike the 1-parameter results. This is seen in bins 5 and 6, which do overlap in redshift, with each having a negative bias compared to the true width. We believe if there is a correlation here though, it is likely not due to covariance of the redshift bins, but a general bias for noisier clustering redshift measurements to predict too wide a redshift distribution (negative bias on Figure \ref{fig:wztests2param}). We did see in tests that adding scales to the measurements, which would decrease the uncertainties notably for higher redshift bins, also reduced this bias on the stretch parameter. We leave further study of this correlation of noise and bias in measured width to future work.

\begin{table}
\small\addtolength{\tabcolsep}{-5pt}
\begin{center}
    \begin{tabular}{|c|c|c|}
      \hline
      \multicolumn{3}{|c|}{Methodology Test Results: \redmagic alone} \\
      \hline
      Name of Test & Bias & Uncertainty \\
      \hline
      Test 1: Method w/Spec-$z$ & -0.0019 & 0.0033 \\  
      \hline
      Test 5: Clustering-$z$ $\sigma_j$ correction & 0.0026 & 0.0029\\   
      \hline
      \hline
      \multicolumn{3}{|c|}{Methodology Test Results: \maglim alone} \\
      \hline
      Name of Test & Bias & Uncertainty \\
      \hline
      Test 1: Method w/Spec-$z$ & -0.0007 & 0.0000 \\  
      \hline
      Test 5: Clustering-$z$ $\sigma_j$ correction & 0.0018 & 0.0034\\   
      \hline
    \end{tabular}
  \caption{Re-analysis of the tests shown in Figure \ref{fig:wztests} and Table \ref{table:test_results}. In this case, we derive separate bias and uncertainty values when analyzing the three simulated \redmagic and three simulated \maglim samples separately.}
  \label{table:test_results_appendix}
\end{center}
\end{table}

\section{Tests of Methodology's Dependence on Density}
\label{sec:densitydependenceappendix}

For any clustering statistics, uncertainty will depend significantly on number of objects. In the pair counting of $w(r)$ in Equation \ref{landyszalay}, a lower density of objects will result in larger uncertainty, which we call in this work the statistical uncertainty. We briefly investigate if there is density dependence as well for the systematic errors determined in this section.

To do this, we take our six simulated samples shown in Figure \ref{fig:simwz}, as well as the simulated spectroscopic samples, and cut their regions to approximately 1/4th of their original size. We then recompute each step of our methodology: taking clustering-$z$ measurements with cross-correlations and auto-correlations of the sample. We repeat `Test 1' by taking auto-correlations of the unknown (photometric) samples using their true redshifts. We also repeat `Test 5', the method we will use on the data, where we take the photometrically binned (micro-binned) unknown sample auto-correlations, and use cross-correlations on nano-bins to calibrate those auto-correlations (Equations \ref{stnddev}-\ref{bothautomod}). We finally recompute the 2-parameter tests for the $\chi^2$ method as well. In each case we re-derive the systematic uncertainty or bias and compare to the values from Tables \ref{table:test_results}-\ref{table:test_results_2param}. 

The new fits for systematic uncertainty in this 1/4th simulated data case are shown in Table \ref{table:densitytestresults}. We see that the measures of bias in Tests 1 and 5 of Section \ref{sec:simtests} are the same or reduced, giving evidence that perhaps there is no overall bias of the method. We also see that the systematic uncertainty in both 1-parameter ($\omega$) and 2-parameter tests is increased.

\begin{table}
\small\addtolength{\tabcolsep}{-5pt}
\begin{center}
    \begin{tabular}{|c|c|c|c|c|}
      \hline
      \multicolumn{4}{|c|}{1/4th Simulated Data Test Results} \\
      \hline
      Method & Parameter & New Value & Prev. Value \\
      \hline
      1-Param Test 1 & $b$ & -0.0014 & -0.0014 \\   
      \hline
      1-Param Test 5 & $\omega$ & 0.0022 & 0.0013 \\   
      \hline
      1-Param Test 1 & $b$ & 0.0014 & 0.0021 \\   
      \hline
      1-Param Test 5 & $\omega$ & 0.0038 & 0.0021 \\   
      \hline
      2-Param $\chi^2$ & $\delta \Delta z$ & 0.0044 & 0.0057 \\   
      \hline
      2-Param $\chi^2$ & $\delta s$ & 0.038 & 0.0067 \\
      \hline
    \end{tabular}
  \caption{Results from recomputing the systematic uncertainty parameters of Tables \ref{table:test_results}-\ref{table:test_results_2param} when using simulated samples that are 1/4th the size of the original ones used in Section \ref{sec:methods}.}
  \label{table:densitytestresults}
\end{center}
\end{table}

From this result, it is of interest to extrapolate how systematic uncertainties on the data may be underestimated. In the test described here, both the reference and unknown sample were cut into 1/4th the size. The uncertainties measured in Table \ref{table:densitytestresults} increase by a factor of 1.3-1.8. As a toy model, we will say the uncertainties increased by roughly $4^{1/3}=1.587$. This toy model would suggest systematic uncertainty scales as $\sqrt{N_{\text{r}} N_{\text{u}}}^{1/3}$. The data redshift bins have a range of number density contrasts with the full simulated samples. In this toy model, the systematic uncertainty in different redshift bins for \maglim and \redmagic would range from 2-3 times the uncertainties derived in \ref{sec:simtests}. Since the systematic uncertainty is added in quadrature with the statistical uncertainty, the overall uncertainty would raise by less than a factor of two in most bins.

There are many caveats to this extrapolation. Most notably, if there is significant density dependence to the systematic uncertainty, then averaging the results of different simulated samples of different densities is likely not the optimal analysis. The \cite{y3-3x2ptkp} cosmological analysis ended up using the four most dense redshift bins (\maglim bins 1-4). It is likely in these bins that any increase in systematic uncertainty is on the lower end of this projection. If averaging different densities across the six simulated samples has a large effect, then the increases for these bins may be significantly overestimated here.

We tested the impact of having increased redshift uncertainty on the cosmology analyses in \cite{y3-3x2ptkp} and \cite{y3-2x2ptaltlensresults}. We ran chains where the systematic redshift uncertainties for the 2-parameter $\chi^2$ fit from Table \ref{table:test_results_2param} were increased to Unc.$(\Delta z)=0.01$ and Unc.$(s)=0.1$, which increased the overall uncertainties in Table \ref{table:maglim_2param} by a factor of 2 or less. The main cosmological results used \maglim bins 1-4, so we did not need to consider the 1-parameter fits, which were only used for \redmagic. We ran chains for the `3x2' analysis of \cite{y3-3x2ptkp} and the `2x2' analysis of \cite{y3-2x2ptaltlensresults}. In checking the impact on cosmological parameters $\sigma_8, S_8 \ \text{and} \ \Omega_{\text{m}}$, we found in all cases parameter shifts of less than $0.17 \sigma$ and contour increases of less than $9\%$.

We note that in this Section we are discussing systematic uncertainties beyond the statistical uncertainties, which also change with density. Interestingly, the statistical uncertainties have lower dependence on density. They are within a factor of 1.5 in comparing the simulated samples to data. This is less than the factor of 2-3 from the extrapolation of the systematic uncertainty. This may also suggest more tests are needed to verify the magnitude of systematic uncertainty dependence on density.

The test in this section suggests that the accuracy and precision of clustering redshift methods may be dependent on density, beyond the usual counting statistics. More precise work with a larger suite of simulated samples at a range of densities will be needed to understand these effects in detail. The upper bounds of the increase in uncertainty from this Section though only increase the overall uncertainty by up to a factor of 2, and this has minimal impact on the cosmological results.

Several other parameters not explored here may affect the precision of the clustering redshift measurements as well, including the exact shape of the $n(z)$, the shape of the photo-$z$ estimate, the width of the redshift distribution,  the shape and strength of galaxy bias evolution with redshift, and number of objects. Of these, the most likely correlation is with number of objects, so that is where we investigated further. Future studies will attempt to quantify to higher degree the dependence of the clustering redshift methodology on these several factors, as well as choices like bin size and scales of measurement.

\section{DES Flux-limited Sample}
\label{sec:fluxlim}
In this section, we describe clustering redshift measurements for the DES `flux-limited' sample described in \cite{y3-2x2maglimforecast}. This sample is not used in the DES Year-3 cosmology analyses (\citealt{y3-3x2ptkp}). In \cite{y3-2x2maglimforecast}, this sample's cosmological constraining power is compared to the \redmagic and \maglim samples.

\begin{figure*}
\begin{center}
\includegraphics[width=1.0 \textwidth]{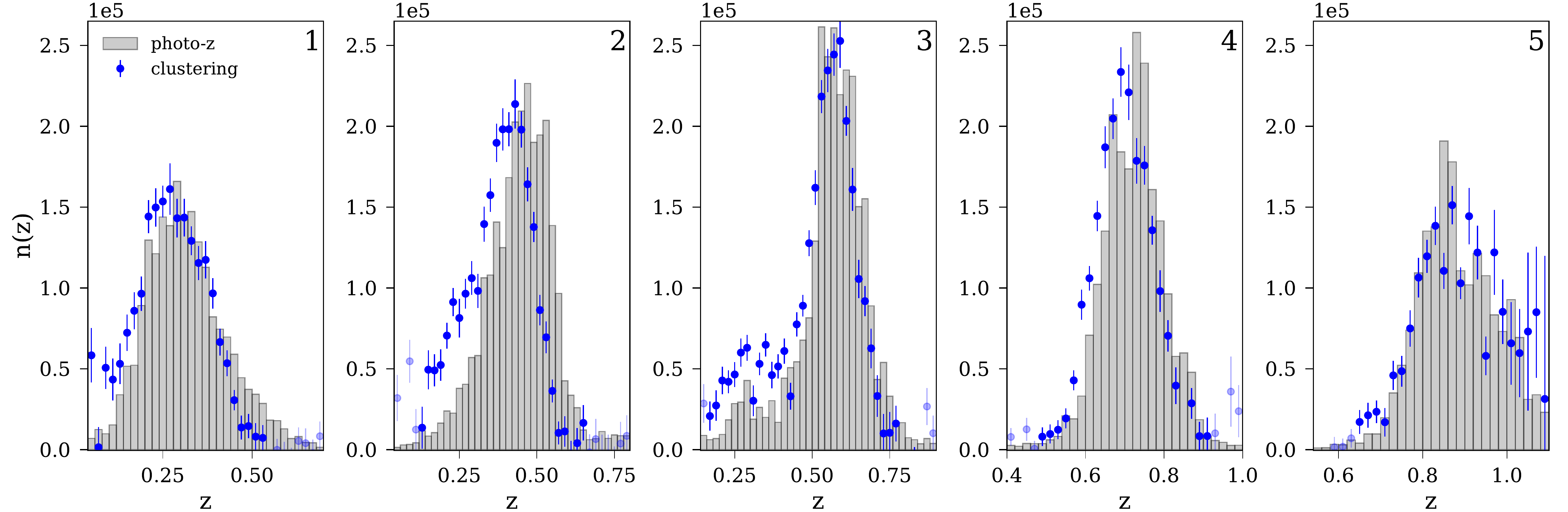}
\end{center}
\caption{The clustering redshift measurements of the flux-limited DES sample. This sample is not used in DES Year-3 cosmology analyses, but was studied in \protect\cite{y3-2x2maglimforecast}.}
\label{fig:fluxlim}
\end{figure*}

We do the same full clustering redshift analysis as was done for the \redmagic and \maglim samples in Section \ref{sec:results}. The results are shown in Figure \ref{fig:fluxlim} and Table \ref{table:fluxlim}. We see that this sample shows considerable photo-$z$ biases of several $\sigma$, in sharp contrast to the results on the other two DES lens samples. Most notable are large excesses of low-$z$ galaxies measured by the clustering redshifts in the first three tomographic bins ($z \in [0.2,0.65]$) compared to the photo-$z$ predictions.

\begin{table}
\begin{center}
    \begin{tabular}{|c|c|c|c|}
      \hline
      \multicolumn{4}{|c|}{Flux-limited Results (1-parameter)} \\
      \hline
      Redshift Bin & $\Delta z$ & $\delta \Delta z(\text{syst.})$ & $\delta \Delta z(\text{stat.})$ \\
      \hline
      1: $z_{\text{ph}} \in [0.2,0.4]$ & -0.041 $\pm$ 0.007 & 0.005 & 0.005  \\   
      \hline
      2: $z_{\text{ph}} \in [0.4,0.5]$ & -0.058 $\pm$ 0.007 & 0.006 & 0.004  \\   
      \hline
      3: $z_{\text{ph}} \in [0.5,0.65]$ & -0.056 $\pm$ 0.007 & 0.005 & 0.005  \\   
      \hline
      4: $z_{\text{ph}} \in [0.65,0.8]$ & -0.026 $\pm$ 0.006 & 0.005 & 0.003  \\   
      \hline
      5: $z_{\text{ph}} \in [0.8,1.05]$ & 0.008 $\pm$ 0.015 & 0.004 & 0.014  \\   
      \hline
    \end{tabular}
  \caption{Clustering redshift results for the Flux-limited sample. This sample is not used in the DES Year-3 cosmology analyses. The systematic uncertainties listed include the power law uncertainty and the 0.0025 method uncertainty from Section \ref{sec:simtests}.}
  \label{table:fluxlim}
\end{center}
\end{table}

\section*{Affiliations}
$^{1}$ Physics Department, 2320 Chamberlin Hall, University of Wisconsin-Madison, Madison, WI  53706, USA\\
$^{2}$ Physics Department, William Jewell College, Liberty, MO 64068, USA\\
$^{3}$ Center for Cosmology and Astro-Particle Physics, The Ohio State University, Columbus, OH 43210, USA\\
$^{4}$ Department of Physics, The Ohio State University, Columbus, OH 43210, USA\\
$^{5}$ Institut d'Estudis Espacials de Catalunya (IEEC), 08034 Barcelona, Spain\\
$^{6}$ Institute of Space Sciences (ICE, CSIC),  Campus UAB, Carrer de Can Magrans, s/n,  08193 Barcelona, Spain\\
$^{7}$ Institut de F\'{\i}sica d'Altes Energies (IFAE), The Barcelona Institute of Science and Technology, Campus UAB, 08193 Bellaterra (Barcelona) Spain\\
$^{8}$ Kavli Institute for Particle Astrophysics \& Cosmology, P. O. Box 2450, Stanford University, Stanford, CA 94305, USA\\
$^{9}$ SLAC National Accelerator Laboratory, Menlo Park, CA 94025, USA\\
$^{10}$ Instituto de Astrofisica de Canarias, E-38205 La Laguna, Tenerife, Spain\\
$^{11}$ Universidad de La Laguna, Dpto. Astrof\'{\i}sica, E-38206 La Laguna, Tenerife, Spain\\
$^{12}$ Department of Astronomy, University of California, Berkeley,  501 Campbell Hall, Berkeley, CA 94720, USA\\
$^{13}$ Santa Cruz Institute for Particle Physics, Santa Cruz, CA 95064, USA\\
$^{14}$ Department of Physics, Duke University Durham, NC 27708, USA\\
$^{15}$ Centro de Investigaciones Energ\'eticas, Medioambientales y Tecnol\'ogicas (CIEMAT), Madrid, Spain\\
$^{16}$ Department of Physics, Stanford University, 382 Via Pueblo Mall, Stanford, CA 94305, USA\\
$^{17}$ Fermi National Accelerator Laboratory, P. O. Box 500, Batavia, IL 60510, USA\\
$^{18}$ Cerro Tololo Inter-American Observatory, NSF's National Optical-Infrared Astronomy Research Laboratory, Casilla 603, La Serena, Chile\\
$^{19}$ Departamento de F\'isica Matem\'atica, Instituto de F\'isica, Universidade de S\~ao Paulo, CP 66318, S\~ao Paulo, SP, 05314-970, Brazil\\
$^{20}$ Laborat\'orio Interinstitucional de e-Astronomia - LIneA, Rua Gal. Jos\'e Cristino 77, Rio de Janeiro, RJ - 20921-400, Brazil\\
$^{21}$ Instituto de Fisica Teorica UAM/CSIC, Universidad Autonoma de Madrid, 28049 Madrid, Spain\\
$^{22}$ Institute of Cosmology and Gravitation, University of Portsmouth, Portsmouth, PO1 3FX, UK\\
$^{23}$ CNRS, UMR 7095, Institut d'Astrophysique de Paris, F-75014, Paris, France\\
$^{24}$ Sorbonne Universit\'es, UPMC Univ Paris 06, UMR 7095, Institut d'Astrophysique de Paris, F-75014, Paris, France\\
$^{25}$ Department of Physics \& Astronomy, University College London, Gower Street, London, WC1E 6BT, UK\\
$^{26}$ Department of Astronomy, University of Illinois at Urbana-Champaign, 1002 W. Green Street, Urbana, IL 61801, USA\\
$^{27}$ National Center for Supercomputing Applications, 1205 West Clark St., Urbana, IL 61801, USA\\
$^{28}$ INAF-Osservatorio Astronomico di Trieste, via G. B. Tiepolo 11, I-34143 Trieste, Italy\\
$^{29}$ Institute for Fundamental Physics of the Universe, Via Beirut 2, 34014 Trieste, Italy\\
$^{30}$ Observat\'orio Nacional, Rua Gal. Jos\'e Cristino 77, Rio de Janeiro, RJ - 20921-400, Brazil\\
$^{31}$ Department of Physics, University of Michigan, Ann Arbor, MI 48109, USA\\
$^{32}$ Department of Physics and Astronomy, University of Utah, 115 S. 1400 E., Salt Lake City, UT 84112, USA\\
$^{33}$ Department of Physics, IIT Hyderabad, Kandi, Telangana 502285, India\\
$^{34}$ Department of Physics and Astronomy, University of Pennsylvania, Philadelphia, PA 19104, USA\\
$^{35}$ Institute of Theoretical Astrophysics, University of Oslo. P.O. Box 1029 Blindern, NO-0315 Oslo, Norway\\
$^{36}$ Kavli Institute for Cosmological Physics, University of Chicago, Chicago, IL 60637, USA\\
$^{37}$ School of Mathematics and Physics, University of Queensland,  Brisbane, QLD 4072, Australia\\
$^{38}$ Center for Astrophysics $\vert$ Harvard \& Smithsonian, 60 Garden Street, Cambridge, MA 02138, USA\\
$^{39}$ Lawrence Berkeley National Laboratory, 1 Cyclotron Road, Berkeley, CA 94720, USA\\
$^{40}$ Institute of Physics, Laboratory of Astrophysics, Ecole Polytechnique F\'ed\'erale de Lausanne (EPFL), Observatoire de Sauverny, 1290 Versoix, Switzerland\\
$^{41}$ Australian Astronomical Optics, Macquarie University, North Ryde, NSW 2113, Australia\\
$^{42}$ Lowell Observatory, 1400 Mars Hill Rd, Flagstaff, AZ 86001, USA\\
$^{43}$ Department of Astrophysical Sciences, Princeton University, Peyton Hall, Princeton, NJ 08544, USA\\
$^{44}$ Instituci\'o Catalana de Recerca i Estudis Avan\c{c}ats, E-08010 Barcelona, Spain\\
$^{45}$ Faculty of Physics, Ludwig-Maximilians-Universit\"at, Scheinerstr. 1, 81679 Munich, Germany\\
$^{46}$ Max Planck Institute for Extraterrestrial Physics, Giessenbachstrasse, 85748 Garching, Germany\\
$^{47}$ Institute of Astronomy, University of Cambridge, Madingley Road, Cambridge CB3 0HA, UK\\
$^{48}$ Waterloo Centre for Astrophysics, Dept. of Physics and Astronomy, University of Waterloo, 200 University Ave. W., Waterloo ON N2L 3G1, Canada\\
$^{49}$ Department of Physics and Astronomy, Sejong University, Seoul 143-747, Korea\\
$^{50}$ School of Physics and Astronomy, University of Southampton,  Southampton, SO17 1BJ, UK\\
$^{51}$ Computer Science and Mathematics Division, Oak Ridge National Laboratory, Oak Ridge, TN 37831\\
$^{52}$ Department of Physics and Astronomy, Pevensey Building, University of Sussex, Brighton, BN1 9QH, UK\\




\bsp	
\label{lastpage}
\end{document}